\documentclass[%
 reprint,
 amsmath,amssymb,
 aps,
]{revtex4-1}
\usepackage{float}
\usepackage{graphicx}% Include figure files
\usepackage{dcolumn}% Align table columns on decimal point
\usepackage{bm}% bold math
\begin{document}

\preprint{a}

\title{Long-lived heavy quarks : a review}% Force line breaks with \\

\author{Mathieu Buchkremer}
\email{mathieu.buchkremer@uclouvain.be}
\affiliation{Centre for Cosmology, Particle Physics and Phenomenology (CP3),\\
Universit\'{e} catholique de Louvain,\\
Chemin du Cyclotron, 2, B-1348, Louvain-la-Neuve, Belgium.}

\author{Alexander Schmidt}
\email{Alexander.Schmidt@cern.ch}
\affiliation{Institut f\"ur Experimentalphysik\\ Universit\"{a}t Hamburg\\ Luruper Chaussee 149\\ 22761 Hamburg, Germany}

%\author{Mathieu Buchkremer}

%\affiliation{Centre for Cosmology, Particle Physics and Phenomenology (CP3)\\
%Universit\'{e} catholique de Louvain\\
%Chemin du Cyclotron 2 \\ B-1348, Louvain-la-Neuve, Belgium \\ email: mathieu.buchkremer@uclouvain.be}%
%\author{Alexander Schmidt}
%\affiliation{Institut f\"ur Experimentalphysik\\ Universit\"{a}t Hamburg\\ Luruper Chaussee 149\\ 22761 %Hamburg, Germany \\ email: Alexander.Schmidt@cern.ch}%

             
%\keywords{Beyond Standard Model, Hadron Colliders, Higgs Physics, Long-lived, Fourth Generation, Vector-Like, Heavy Quarks, Displaced Vertex, Heavy Stable Charged Particles}

\begin{abstract}

We review the theoretical and experimental situation for long-lived heavy quarks, or bound states thereof, arising in simple extensions of the Standard Model. If these particles propagate large distances before their decay, they give rise to specific signatures requiring dedicated analysis methods. In particular, vector-like quarks with negligible couplings to the three known families could have eluded the past experimental searches. While most analyses assume prompt decays at the production vertex, novel heavy quarks might lead to signatures involving displaced vertices, new hadronic bound states, or decays happening outside of the detector acceptance. We perform reinterpretations of existing searches for short- and long-lived particles, and give suggestions on how to extend their reach to long-lived heavy quarks.
%\begin{description}
%\end{description}
\end{abstract}

\maketitle

\preprint{a}

% Force line breaks with \\

\affiliation{Centre for Cosmology, Particle Physics and Phenomenology (CP3)\\
Universit\'{e} catholique de Louvain\\
Chemin du Cyclotron 2 \\ B-1348, Louvain-la-Neuve, Belgium \\ email: mathieu.buchkremer@uclouvain.be}

%\collaboration{MUSO %Collaboration}%\noaffiliation

\affiliation{Institut f\"ur Experimentalphysik\\ Universit\"{a}t Hamburg\\
Luruper Chaussee 149\\ 22761 Hamburg, Germany \\ email:
Alexander.Schmidt@cern.ch}



%\tableofcontents

%\author[a]{Mathieu Buchkremer}
%\affiliation[a]{Centre for Cosmology, Particle Physics and Phenomenology (CP3),\\
%Universit\'{e} catholique de Louvain,\\
%Chemin du Cyclotron, 2, B-1348, Louvain-la-Neuve, Belgium.}
%\email[a]{mathieu.buchkremer@uclouvain.be}
%\author[b]{Alexander Schmidt}
%\affiliation[b]{Institut f\"ur Experimentalphysik\\ Universit\"{a}t Hamburg\\ Luruper Chaussee 149\\ %22761 Hamburg, Germany.}
%\email[b]{Alexander.Schmidt@cern.ch}

\section{Introduction}
Over the past decades, we have discovered that Nature consists of a given number of elementary particles, deeply connected with the known fundamental forces driving our universe. The recent observation of a new particle resembling the long-sought Higgs boson at the Large Hadron Collider (LHC) now provides us with strong evidence for the validity of the Standard Model (SM) \cite{:2012gk,:2012gu}. Yet, while it is generally acknowledged that the latter comprises three generations of chiral quarks and leptons, various fundamental problems do not find their answers within this framework. For instance, long-standing issues such as the origin of the fermion mass hierarchy or the nature of the Cabibbo-Kobayashi-Maskawa (CKM) mixing matrix hint at the possible need for new physics. While the SM is expected to loose its predictive power as the experimental searches now reach the TeV scale, models suggesting new heavy fermions might offer solutions to these problems in the near future.

Although a fourth sequential family of quarks is now disfavoured by the recent results of the searches for the Higgs boson  \cite{Eberhardt:2012sb}, other models predicting new heavy particles beyond the top quark are still consistent with the current experimental measurements. In particular, the possibility for vector-like quarks, \textit{i.e.}, quarks having their left- and right- handed components transforming identically under the electroweak gauge group, are a common feature in many scenarios going beyond the SM, e.g. extra dimensional models, Little Higgs models, grand unified theories, and so forth \cite{Frampton:1999xi}. Non-chiral quarks also have the peculiarity to decouple in the heavy mass limit, leading to SM-like signals. Should the mixings of these with the light SM fermions be suppressed, the Higgs production rates would not be easily distinguishable from the SM expectations \cite{Dawson:2012di}. These new quarks could be sufficiently long-lived to behave like effectively stable particles, evading the current searches as they propagate over sizeable distances. 

From the collider searches it is clear that if new states with masses less than a hundred GeV had existed, they would have been observed. On the other hand, the experimental reach for detecting novel heavy stable particles above the TeV scale depends on how readily identifiable such states are at the LHC. While particles with nanoseconds lifetimes could fly away from the primary interaction point before they decay to ordinary particles and lead to displaced vertices, they can also hadronise, allowing for a possibly rich spectrum of new exotic and heavy bound states. Such stable massive particles are anticipated in many new physics models, either due to the presence of a new conserved quantum number (\textit{e.g.}, $R$ parity in supersymmetric models), or because the decays are suppressed by kinematics or small couplings (see \cite{Fairbairn:2006gg} and references therein for an exhaustive review).
Although the majority of the past studies focused on stable states in supersymmetry, the possibility for new stable quarks received few attention at the LHC. While the CMS and ATLAS experiments are now setting strong limits on long-lived gluino, stop and stau pair production (see, for instance, \cite{:2012yg} and \cite{Aad:2011hz}), scenarios involving quasi-stable heavy quarks have been barely investigated at the LHC. 
\newpage
In the following, we review the phenomenology of new, long-lived heavy quarks. Assuming a general parametrisation, Section \ref{sec:longLivedSig} first examines their production and decay modes at the LHC, considering both the chiral and vector-like scenarios. The possible signatures for displaced vertices and stable massive hadrons are then covered within the context of negligible mixings with the SM fermions. The experimental aspects of the collider searches for new long-lived quarks are presented in Section \ref{sec:experiment}. We review the situation for prompt decay searches, displaced vertices and Heavy Stable Charged Particles (hereafter, HSCPs) and highlight some of their limitations. Reinterpretations of the existing searches are then presented for short- and long-lived particles. Considering the possibility to improve the sensitivity of the existing analyses to long-lived quarks beyond the TeV scale, some alternative search topologies are finally described. Our conclusions are given in Section \ref{sec:conclusion}. 

\section{Long-lived quarks signatures \label{sec:longLivedSig}}
\subsection{Production \label{sec:prod}}

As we review the possibility for long-lived quarks in general, two different scenarios will be distinguished in this work. New chiral fermions, for which the left- and right-handed chiralities have different charges under the electroweak gauge group, gain their masses from the electroweak symmetry breaking mechanism. If heavy, a sequential fourth generation quark doublet $(t^{\prime},b^{\prime})_{L}$ can therefore induce large corrections to loop observables. Such a scenario is now severely constrained from the electroweak precision data and the Higgs searches results. Given that chiral fermions couple to the Higgs boson with a strength proportional to their Yukawa couplings, they do not decouple from its production, as the corresponding rate should increase by a factor of about 9 due to the additional fermion loops occurring in gluon fusion. The unobserved enhancement thus now strongly disfavours a fourth generation Standard Model \cite{Eberhardt:2012sb}.

Non-chiral quarks, on the other hand, still provide a viable extension to the SM and certainly require a dedicated discussion. Considering the introduction of new vector-like heavy partners mixing with the SM quarks, additional parameters are allowed from $SU(2)_{L}\times U(1)_{Y}$ invariant Yukawa interactions and Dirac mass terms in the SM Lagrangian. The light quark couplings to the Higgs and electroweak bosons are consequently modified, along with the introduction of new couplings between the heavy and the SM quarks. 

Still, it is important to emphasise that such new states cannot have arbitrary quantum numbers, since they can only mix with the SM quarks through a limited number of gauge-invariant couplings. Classifying them into $SU(2)_{L}$ multiplets, their Yukawa terms only allow for seven distinct possibilities, \emph{i.e.}, the two singlets, the three doublets and the two triplets displayed in Table \ref{tab:multiplets}.
\begin{table}[t]
\begin{equation*}
\begin{tabular}{c|ccccccc}
\hline\hline
$Q_{q}$ & $T_{\frac{2}{3}}$ & $B_{-\frac{1}{3}}$ & $%
\begin{pmatrix}
X_{\frac{5}{3}} \\ 
T_{\frac{2}{3}}%
\end{pmatrix}%
$ & $%
\begin{pmatrix}
T_{\frac{2}{3}} \\ 
B_{-\frac{1}{3}}%
\end{pmatrix}%
$ & $%
\begin{pmatrix}
B_{-\frac{1}{3}} \\ 
Y_{-\frac{4}{3}}%
\end{pmatrix}%
$ & $%
\begin{pmatrix}
X_{\frac{5}{3}} \\ 
T_{\frac{2}{3}} \\ 
B_{-\frac{1}{3}}%
\end{pmatrix}%
$ & $%
\begin{pmatrix}
T_{\frac{2}{3}} \\ 
B_{-\frac{1}{3}} \\ 
Y_{-\frac{4}{3}}%
\end{pmatrix}%
$ \\ \hline
$T_{3}$ & $0$ & $0$ & $1/2$ & $1/2$ & $1/2$ & $1$ & $1$ \\ \hline
$Y$ & $2/3$ & $-1/3$ & $7/6$ & $1/6$ & $-5/6$ & $2/3$ & $-1/3$ \\ 
\hline\hline
\end{tabular}%
\end{equation*}
\caption{Vector-like multiplets allowed to mix with the SM quarks through Yukawa couplings. The
electric charge is the sum of the third component of isospin $T_{3}$ and of the hypercharge $Y$. Adapted from \cite{delAguila:2000rc}. }
\label{tab:multiplets}
\end{table}

The production of such new heavy quarks, either chiral or vector-like, is usually assumed to proceed dominantly at the LHC through gluon fusion, $gg\rightarrow Q\bar{Q}$. Nevertheless, depending on the model at hand, electroweak single production can also provide an alternative mechanism as it is not affected
by the large phase-space suppression from which pair production suffers. In particular, new heavy quarks can be produced singly in flavour-changing processes via the electroweak interaction through $q_{i} \overset{\textbf{\fontsize{3pt}{3pt}\selectfont(---)}}{q_{j}} \rightarrow V^{\ast} \rightarrow q_{k} Q$,  where $V=W,Z$. A comparison between the dominant production modes can be found in \cite{Campbell:2009gj,Buchkremer:2012yy} for fourth generation, and in \cite{Atre:2011ae} for vector-like quarks. Assuming order unity couplings, benchmark cross-sections have been evaluated for various mass scenarios and reproduced in Table \ref{tab:xsections}.

\begin{table}[b]
\begin{equation*}
\begin{tabular}{ccc}
\hline\hline
channel & $\sqrt{s}=7\ $TeV & $\sqrt{s}=14\ $TeV \\ 
& $m_{Q}=900\ $GeV/$c^2$ & $m_{Q}=1800\ $GeV/$c^2$ \\ \hline
$pp\rightarrow W^{\ast }\rightarrow jX$ & 1.4 & 0.36 \\ 
$pp\rightarrow W^{\ast }\rightarrow j\bar{X}$ & 0.037 & 0.0092 \\ 
$pp\rightarrow W^{\ast }\rightarrow jT$ & 0.61 & 0.16 \\ 
$pp\rightarrow W^{\ast }\rightarrow j\bar{T}$ & 0.052 & 0.013 \\ 
$pp\rightarrow Z^{\ast }\rightarrow jT$ & 0.43 & 0.11 \\ 
$pp\rightarrow Z^{\ast }\rightarrow j\bar{T}$ & 0.025 & 0.0064 \\ 
$pp\rightarrow W^{\ast }\rightarrow jB$ & 0.69 & 0.18 \\ 
$pp\rightarrow W^{\ast }\rightarrow j\bar{B}$ & 0.089 & 0.022 \\ 
$pp\rightarrow Z^{\ast }\rightarrow jB$ & 0.18 & 0.047 \\ 
$pp\rightarrow Z^{\ast }\rightarrow j\bar{B}$ & 0.034 & 0.0088 \\ 
$pp\rightarrow W^{\ast }\rightarrow jY$ & 0.29 & 0.074 \\ 
$pp\rightarrow W^{\ast }\rightarrow j\bar{Y}$ & 0.12 & 0.031 \\ \hline\hline
\end{tabular}%
\end{equation*}
\caption{Electroweak single production cross-sections (in pb) as a function of the mass $m_{Q}$ of the heavy quark, assuming order unity couplings as computed in \cite{Atre:2011ae}. In contrast, strong pair production $pp\rightarrow Q\bar{Q}$ yields $3.61$ ($0.63$) fb for $m_{Q}=900$ ($1800$) GeV/$c^2$ at $\sqrt{s}=7$ $(14)$ TeV. }
\label{tab:xsections}
\end{table}

As described in \cite{Atre:2011ae}, the leading contributions to vector-like $T$ and $B$ single production arise from $ud\rightarrow W^{\ast }\rightarrow jT$, $du\rightarrow W^{\ast }\rightarrow jB$, $qd\rightarrow Z^{\ast }\rightarrow jB$ and $qu\rightarrow Z^{\ast }\rightarrow jT$, where $q$ denotes a valence quark parton and $j$ a generic light quark jet. Unlike QCD pair production, the above processes however scale with the $Q-q$ quark coupling squared, and can be strongly suppressed if the associated mixings are negligibly small. On the other hand, Figs. $2-4$ in \cite{Atre:2011ae} indicate that $\sigma(pp\rightarrow Q\bar{Q})$ falls off faster than the single production cross-sections as soon as $m_{Q}$ reaches the TeV scale, while the electroweak production channels displayed in Table \ref{tab:xsections} are enhanced by a factor $O(m_{Q}^{2}/m_{V}^{2})$, originating from the longitudinal polarisation of the gauge boson $V$. Furthermore, the relative rates of the $jQ$ and $j\bar{Q}$ channels strongly depend on the valence quark density in the initial state. Given the difference in the PDFs of valence and sea quarks in the initial states, new heavy quarks are produced singly at a much higher rate than antiquarks for increasing $m_{Q}$. While the $u$ quark contributes more than the $d$ quark, the neutral current contributions are weaker than the charged current ones. Finally, we underline the absence of the tree-level processes $pp\rightarrow Z^{\ast }\rightarrow jQ$ for vector-like quarks involving exotic charges, given that they only interact with other states via tree-level charged currents \cite{Atre:2011ae}. Although most of the above statements are model-dependent, we conclude that the large running energy of the LHC makes single electroweak production a promising discovery channel when considering new heavy quarks searches with $m_{Q}\gtrsim 1$ TeV/$c^2$.

\subsection{Decay \label{sec:dec}}

In the search of new fermions, the associated decay modes
require careful attention. The exact partial width for a new sequential heavy
quark $Q$ decaying on-shell to a light quark $q$ through a charged current
can be written as%
\begin{widetext}
\begin{equation}
\Gamma (Q \rightarrow qW)=\frac{G_{F} m_{Q}^{3}}{8\pi \sqrt{2}}%
\text{ }|\kappa_{Qq}|^{2}\text{ }f_{2}(\frac{m_{q}}{m_{Q}},\frac{m_{W}}{m_{Q}}),  \label{W1}
\\
\end{equation}%
\begin{equation}
f_{2}(\alpha ,\beta ) =[(1-\alpha ^{2})^{2}+\beta ^{2}(1+\alpha
^{2})-2\beta ^{4}]\text{ }\sqrt{[1-(\alpha +\beta )^{2}][1-(\alpha -\beta )^{2}]}, \label{W2}
\end{equation}%
\end{widetext}
where $\kappa_{Qq}$ denotes the generic $Q-q$ quark coupling, equal to the CKM\ mixing matrix element $V_{Qq}$ when considering a new sequential family of quarks. Interestingly, the width (\ref{W1})
holds for both chiral and vector-like singlet quarks, given that all
heavy-to-light charged current quark decays occur to be pure $V-A$ processes
with identical rates \cite{Frampton:1997up}. If $m_{Q}\gg m_{q}$, the above
partial width can be shown to reach the asymptotic form 
\begin{equation}
\Gamma (Q\rightarrow qW)\simeq 170.5\text{ MeV}\times |\kappa_{Qq}|^{2}\times%
\frac{m_{Q}^{3}}{m_{W}^{3}}.  \label{CC}
\end{equation}%
Considering a new heavy quark $Q$ decaying exclusively via the above 2-body decay mode, we evaluate in Table \ref{tab:partial} the corresponding decay lengths, with $c\tau \simeq2\times 10^{-10}$ GeV$/\Gamma ($GeV) $\mu $m and $|\kappa_{Qq}|=1$. With increasing mass, the probability for a new heavy quark $Q$ to decay weakly thus becomes more and more important as its lifetime decreases. However, small couplings to the lighter SM quarks strongly affect this statement as the corresponding widths are suppressed by $|\kappa_{Qq}|^{2}$. As we will see in Section \ref{sec:HSCPsignatures}, this could lead to a very different phenomenology as the quarks $Q$, if very heavy, would form bound states and possibly decay hadronically.
\begin{table}[H]
\begin{tabular}{ccc}
\hline\hline
$m_{Q}$ (GeV/$c^2$) & $\Gamma $ (GeV) & $c\tau $ $\times 10^{12}$ ($\mu $m) \\ 
\hline
300 & 8.73 (2.39) & 22.91 (83.85) \\
400 & 20.90 (11.14) & 9.57 (17.96) \\ 
500 & 40.94 (27.86) & 4.89 (7.18) \\ 
600 & 70.81 (54.52) & 2.82 (3.67) \\ 
700 & 112.50 (93.06) & 1.78 (2.15) \\ 
800 & 167.97 (147.43) & 1.19 (1.38) \\
900 & 239.18 (213.57) & 83.62 (9.36)
$\times 10^{-1}$ \\  
1000 & 328.12 (299.46) & 6.10 (6.68) $\times 10^{-1}$ \\ 
1100 & 436.74 (405.05) & 4.58 (4.94) $\times 10^{-1}$ \\ 
1200 & 567.02 (532.31) & 3.53 (3.76) $\times 10^{-1}$ \\
1300 & 720.93 (683.21) & 2.77 (2.92) $\times 10^{-1}$ \\
1400 & 900.43 (859.72) & 2.22 (2.33) $\times 10^{-1}$ \\
1500 & 1107.5 (1063.79) & 1.81 (1.88) $\times 10^{-1}$ \\
1600 & 1344.1 (1297.4) & 1.49 (1.54) $\times 10^{-1}$ \\
1700 & 1612.2 (1562.5) & 1.24 (1.28) $\times 10^{-1}$ \\
1800 & 1913.8 (1861.1) & 1.05 (1.07) $\times 10^{-1}$ \\
 \hline\hline
\end{tabular}%
\caption{Partial widths and decay lengths for a new heavy
quark $Q$ decaying to light quarks via on-shell charged currents $%
Q\rightarrow qW$, with $m_{q}=0$ (173.2) GeV/$c^2$ and $|\kappa_{Qq}|=1$, as computed
from Eq. (\ref{W1}).}
\label{tab:partial}
\end{table}

Several conclusions can already be drawn at this stage. As far as a heavy fourth quark family $(t^{\prime
},b^{\prime })_{L}$ is concerned, we highlight that the ratios of the
charged current decay rates to the lighter families only depend on the off-diagonal mixing
elements. Assuming that the extended $4 \times 4$ CKM matrix is unitary, the
CKM mixing elements $|V_{Qq}|$ imply non-unity branching ratios for fourth generation quarks in
general.  

The possibility for tree-level Flavour Changing Neutral Currents lead to similar conclusions for models involving new vector-like quarks interacting with the SM families. Depending on their Yukawa couplings, they can undergo tree-level transitions to lighter quarks and $W$, $Z$ and Higgs bosons. Considering vector-like quarks $Q=T,B$ with $m_{Q}\gg m_{q}$, the neutral currents $Q\rightarrow qZ$ and $Q\rightarrow qH$ can occur with a rate of the same magnitude as given by (\ref{CC}), if they are allowed to mix with the light SM quarks $q$. If no assumption is made on the hierarchy of the $Q-q$ coupling strengths, no exclusive decay mode to the first, second or third family should be preferred.

For non-chiral quarks decaying to $Wq$, $Zq$ and $Hq$, long lifetimes can be expected in the $|\kappa_{Qq}|\simeq 0$ limit. Additionally, it is interesting to notice that $\Gamma (Q \rightarrow qZ) / \Gamma (Q\rightarrow qH)\simeq 1$ holds in the large mass limit $m_{Q}\gg m_{q}$ for most scenarios, while the charged-to-neutral current ratios $\Gamma (Q \rightarrow qW) / \Gamma (Q\rightarrow q(Z,H))$ approach either $2$ or $0$ depending on the vector-like representation \cite{Cacciapaglia:2010vn}.  

If more than one new heavy quark is considered, and if they are allowed to decay into each other, the partial width (\ref{CC}) becomes inaccurate due to the possible competition with the heavy-to-heavy transitions $Q_{1}\rightarrow Q_{2}V^{(\ast )}$, where $V=W,Z,H$, depending on the model at hand. Indeed, if the decay rates of the daughter particles are significant with respect to the mass of the parent, the decay of an unstable particle is allowed through threshold effects, even if kinematically forbidden. For both chiral and vector-like quarks, real and virtual $V$\ emission near threshold proceed with the rate 
\begin{widetext}
\begin{equation}
\Gamma (Q_{1} \rightarrow Q_{2}V^{(\ast )})=\frac{G_{F}^{2}m_{Q_{1}}^{5}}{%
192\pi ^{3}}\text{ }|\kappa_{Q_{1}Q_{2}}|^{2}\text{ }f_{3}\Big(\frac{m_{Q_{1}}^{2}}{m_{V}^{2}}%
,\frac{m_{Q_{2}}^{2}}{m_{V}^{2}},\frac{\Gamma _{V}^{2}}{m_{V}^{2}}\Big), \label{3body}
\end{equation}
\begin{equation}
f_{3}(\alpha ,\beta ,\gamma ) =2\int_{0}^{(1-\sqrt{\beta })^{2}}dx\text{ }%
\frac{[(1-\beta )^{2}+x\text{ }(1+\beta )-2x^{2}]\sqrt{\lambda (1,x,\beta )}%
}{[(1-\alpha x)^{2}+\gamma ^{2}]},  
\end{equation}
\end{widetext}
where $\lambda (a,b,c)=a^{2}+b^{2}+c^{2}-2ab-2bc-2ac$ \cite{Frampton:1999xi}. While the $Q\rightarrow qV$ rates essentially depend on the $Q-q$ quark couplings, the heavy quark mass splittings $|m_{Q_{1}}-m_{Q_{2}}|$ drive the strength of the heavy-to-heavy transitions \cite{Chao:2011th,Buchkremer:2012yy}. If the quarks $Q_{1}$ and $Q_{2}$ belong to the same weak multiplet with a large mass splitting, the $Q_{1}\rightarrow Q_{2} V$ decay could be very rapid and dominate the other decay modes if there is no $|\kappa_{Q_{1}Q_{2}}|$ suppression. On the other hand, the lightest partner can only decay into a SM family quark, and is thus plausibly long-lived for small mixings. Should this be the case, we will see in Sections \ref{sec:HSCPsignatures} and \ref{sec:OPENsignatures} that the latter could hadronise, leading to a different decay phenomenology. This occurs specifically for $SU(2)_{L}$ vector-like singlets, which free quark decays can only proceed via non-zero mixing into the lighter generations. 

As far as the doublet and triplet representations are concerned, new vector-like partners should be very close in mass if one assumes that isospin conservation is respected. Although it may be broken by higher order corrections, such a degeneracy has been shown in \cite{Sher:1995tc} to require a mass difference between $70$ and $110$ MeV for non-chiral doublets, allowing for nanoseconds lifetimes. If their couplings to the lighter quarks are negligible, the partial widths for  $X_{5/3}$, $T_{2/3}$, $B_{-1/3}$ and $Y_{-4/3}$ quarks are thus pausibly small. 

Summarising these results, the range of the
observable couplings to which long-lived particles searches are sensitive at
the LHC can be evaluated. From the approximation (\ref{CC}), the partial decay length for a vector-like quark reads \begin{equation}
\lambda \simeq 2\text{ cm }\times\Big(\frac{10^{-8}}{|\kappa_{Qq}|}\Big)^{2} \Big(\frac{500\text{ GeV/$c^2$}}{m_{Q}%
} \Big)^{3}. \label{lambda}
\end{equation}
Assuming $m_Q\lesssim 1$ TeV/$c^2$ and an experimental resolution of at least 25 $\mu$m, the heavy quark decays are prompt only if their associated couplings are above the $10^{-7}$ level.\ If $|\kappa_{Qq}|\lesssim 10^{-7}$, the corresponding decay lengths could be observed over a few tens of centimeters. In the limit of such small mixings with the SM\ fermions, it is thus natural to motivate searches for displaced vertices.

\subsection{Classification of signatures \label{sec:sigDisplacedVert}}

We now detail the possible situations giving rise to the observable signatures of long-lived quarks at the LHC. 

The classification given in Table \ref{tab:decaySignatures} summarises the long-lived quark scenarios corresponding to the small mixing scenarios treated in this work. We emphasise that if mixing with all SM quark families is permitted, direct searches should aim at being sensitive to all light quarks $q$ in the observed final states. Given that their interactions are allowed through arbitrary Yukawa couplings, the searches for long-lived vector-like quarks should be as inclusive as possible in order to cover the full spectrum of possibilities at the LHC. Such a program is important to set new constraints on the $Q\rightarrow qV^{(\ast)}$ decay modes, independently of any assumptions on the $Q-q$ mixing sector. The relative importance of the $Q\rightarrow (u,c,d,s)V$ channels is most relevant compared to the $Q\rightarrow (t,b)V$ charged currents, as no exclusive decay mode should be preferred for $V=W,Z,H$.

\begin{table}[H]
\[
\begin{tabular}{c|ccc}
\hline\hline
& $(i)$ & $(ii)$ & $(iii)$ \\ 
& Short-lived & Intermediate & Long-lived \\ \hline
$|\kappa _{Qq}|$ & $\gtrsim 10^{-7}$ & $\left. 
\begin{array}{c}
\lesssim 10^{-7} \\ 
\gtrsim 10^{-9}%
\end{array}%
\right. $ & $\lesssim 10^{-9}$ \\ 
\hline
$\Gamma $ (GeV) & $\gtrsim 10^{-12}$ & $\left. 
\begin{array}{c}
\lesssim 10^{-12} \\ 
\gtrsim 10^{-16}%
\end{array}%
\right. $ & $\lesssim 10^{-16}$ \\ 
\hline
$\tau $ (s) & $\lesssim 10^{-13}$ & $\left. 
\begin{array}{c}
\gtrsim 10^{-13} \\ 
\lesssim 10^{-9}%
\end{array}%
\right. $ & $\gtrsim 10^{-9}$ \\ \hline\hline
\end{tabular}%
\]
\caption{Possible decay signatures for new long-lived quarks ($m_{Q}=1$ TeV/$c^2$), as discussed in the text.}
\label{tab:decaySignatures}
\end{table}
\begin{figure}[h]
\centering
\includegraphics[scale=0.29]{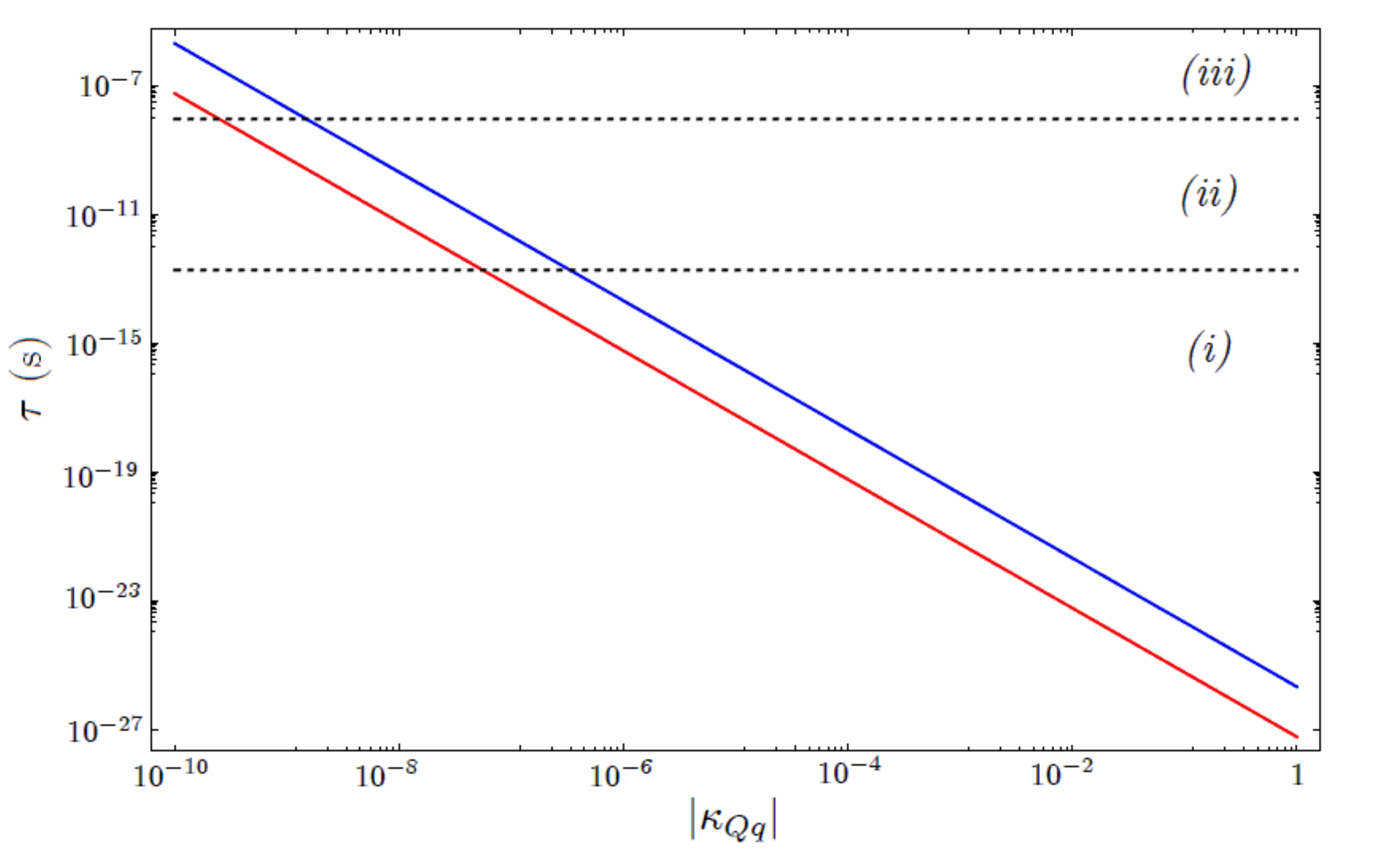}
\caption{Lifetime $\tau $ in seconds ($\gamma \beta \simeq 1$) for a sequential heavy quark $Q$ decaying to the lighter families by emitting a real $W$ boson, for $m_{Q}=300\ $GeV/$c^2$ (in blue) and $m_{Q}=1800$ GeV/$c^2$\ (in red). The three regions of interest refer to the conventions given in Table \ref{tab:decaySignatures}.}
\label{ctau}
\end{figure}

As illustrated in Figure \ref{ctau}, three cases for new long-lived quarks can be distinguished depending on their mixing parameters with the SM fermions : 
\begin{itemize}
\item Firstly, if all couplings with the known SM\ fermions are larger than the $10^{-7}$ level, all heavy-to-light decays $Q \rightarrow qV$ are prompt, over distances smaller than 25 $\mu$m. This defines the short-lived scenario \emph{(i)}, with no experimental difference compared to the direct searches. 
\item The second scenario \emph{(ii)} arises for intermediate decay lengths ranging between a few microns and centimetric distances. If there is at least one light quark $q$ such that $10^{-7} \lesssim |\kappa_{Qq}| \lesssim 10^{-9}$, displaced events could be observed, yet with a partial width suppressed by $|\kappa_{Qq}|^{2}$. Consequently, if all the couplings but one are below the $10^{-7}$ level, a single exclusive channel would lead to a prompt decay signature corresponding to a short-lived heavy quark. A possible exception, though, is the case for new heavy multiplets with sizeable mass splittings, as allowed in extra generation models and possible extensions \cite{Buchkremer:2012yy,Dighe:2012dz,BarShalom:2011zj,Asilar:2011bn}. If $m_{Q_{1}}\gtrsim m_{Q_{2}}+m_{V}$, the heaviest quark $Q_{1}$ can be short-lived and decay semi-weakly to $Q_{2}V^{(\ast )}$, while the lightest partner is likely stable if all its decay modes suffer severe suppression.\ If $m_{Q_{1}}\simeq m_{Q_{2}}$, on the other hand, all heavy-to-heavy transitions are suppressed and both quarks could be long-lived on detector scales. 
\item For particles with decay lengths larger than the detector dimensions lies the long-lived region \emph{(iii)}, the details of which will be discussed in Section \ref{sec:HSCPsignatures}. If all heavy quarks couplings with the SM fermions are below the $10^{-9}$ level, the stable case becomes a relevant scenario, possibly in conjunction with displaced events.
\end{itemize}
We emphasise that the scenarios for which prompt decays and displaced vertices could be observed only call for one of the allowed decay modes to fulfill the associated conditions. The stable scenario, however, requires all decay lengths and partial widths to lie in the ranges given in Table \ref{tab:decaySignatures}. As we will see in Section \ref{sec:HSCPsignatures}, if all $Q-q$ quark couplings to the SM families are smaller than $10^{-2}$, such new heavy fermions could hadronise. As a result, the annihilation decays and hadronic transitions between the formed bound states would dominate, while the single quark decay events might be suppressed.

\subsection{Signatures for heavy stable particles : heavy quarkonia \label{sec:HSCPsignatures}}
Previously, we have discussed how displaced events could occur in the single quark decays
transitions in the case of small $Q-q$ couplings. In this section, we describe the possibility for new heavy quarks to bind into hadrons if their lifetime is large enough, providing us with an interesting alternative for new signal searches at the LHC. 

While new coloured particles with masses heavier than the top quark are usually assumed to decay as free particles, their possible formation into baryons and mesons is an important issue to consider in the case of small couplings with the SM fermions. 

\begin{figure}[h]
\centering\includegraphics[scale=0.62]{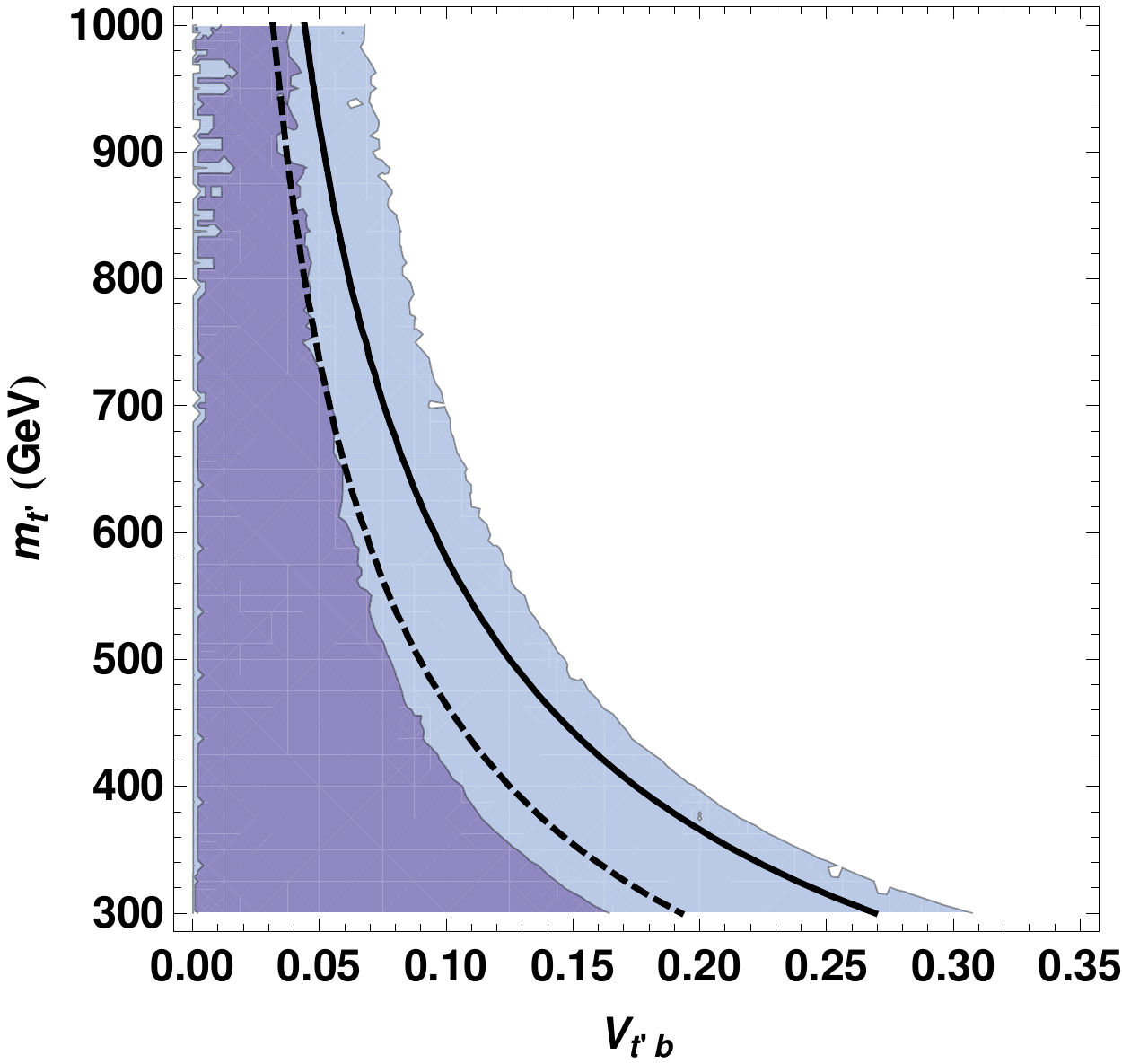}
\caption{Reproduction of the $\protect\chi ^{2}$ fit of
the oblique parameters $S$ and $T$ as given from Fig.\ 2\ in \cite{Buchkremer:2012yy}, considering a chiral fourth generation quark $t^{\prime}$. The upper limits at 68\% (darker blue) and 95\% (lighter blue) CL on the $t^{\prime }$ mass are depicted as a function of the CKM mixing $|V_{t^{\prime }b}|$, freely varied over [300,800] GeV/$c^2$ and [0,0.3], respectively. The solid (dashed) curve denotes the upper bound $m_{t^{\prime}}(|V_{t^{\prime }b}|)$ under which $t^{\prime }\bar{t}^{\prime }$ and $t^{\prime }\bar{q}$ ($\bar{t^{\prime }}q$) bound states can form, as defined by relation (\protect\ref{Bigi}).}
\label{STPlot}
\end{figure}

Should the partial widths $Q \rightarrow q(W,Z,H)$ be suppressed, the quark $Q$ might be stable and hadronise. Assuming that the binding force in a fourth generation bound state is of Coulombic nature, the seminal condition 
\begin{equation}
m_{Q}<125 \text{ GeV} (100 \text{ GeV}) \times |V_{Qq}|^{-2/3}  \label{Bigi}
\end{equation}%
was derived in \cite{Bigi:1986jk} for $Q\bar{Q}$ ($Q\bar{q}$) formation. According to Eq. (\ref{Bigi}), new heavy quarks form hadronic states if their mixing with known quarks is sufficiently small. Mixings roughly smaller than 0.2 are typically required for $300$ GeV/$c^2$ new quarks to form bound states, which remains perfectly consistent with the current bounds from the electroweak precision observables. As shown on Figure \ref{STPlot} for a new up-like fourth generation quark $t^{\prime }$, the electroweak precision fits typically restrict the mixing matrix element $|V_{t^{\prime}b}| \simeq |V_{tb^{\prime}}|$ to be smaller than $10^{-1}$ for $m_{t^{\prime}}>700$ GeV/$c^2$. If lighter, the $t^{\prime}$ total width can be smaller than $10$ MeV \cite{Buchkremer:2012yy}, thus allowing new chiral quarks to form bound states with nanosecond lifetimes. 

As far as non-chiral fermions are concerned, the electroweak precision constraints provide upper bounds on the mixing parameters, but are not as restrictive given that their effects decouple in the limit of large masses \cite{Lavoura:1992np}. Although vector-like quarks are usually considered to couple dominantly to the heaviest third quark family (see \cite{Atre:2011ae,Agashe:2006at} for some exceptions), they are allowed, in principle, to mix with all up- (down-) type quarks \cite{Cacciapaglia:2010vn}. 

In any case, $|\kappa_{Qq}| \lesssim 10^{-2}$ is used as a rule of thumb in most of the allowed vector-like representations \cite{Cacciapaglia:2010vn}. Furthermore, corrections to the CKM\ matrix elements are required to be small for $SU(2)_{L}$ singlets \cite{Barger:1995dd}, doublets and triplets \cite{Yoshikawa:1995et}. 
In this framework, the hadronic production of new bound states involving sequential or non-chiral quarks is a conceivable possibility at the LHC. Assuming that (\protect\ref{Bigi}) is satisfied due to small $Q-q$ couplings, their lifetime can be longer than the orbital period of the corresponding $Q\bar{Q}$ ($Q\bar{q}$) bound state, allowing to build up new quarkonium resonances, open-flavour mesons and baryons \cite{Bigi:1986jk}. 

If their formation is governed by the same strong interactions that are responsible for the existence of the ordinary hadronic spectroscopy, the associated mass spectrum can be determined from the known properties of low-energy hadron physics \cite{Mackeprang:2009ad}. Considering a simplified model of QCD-like hadrons, new heavy quarks would form $Q\bar{Q}$ bound states (hereafter defined as $\eta _{Q}$) from induced, strongly attractive potentials. Conjointly, quarkonium formation can arise from Higgs boson exchange, as the corrections from Yukawa-type forces cannot be neglected for large quark masses \cite{Ishiwata:2011ny,Flambaum:2011ws}. In \cite{Hung:2009hy}, Hung detailed the numerical resolution of a general non-relativistic Higgs exchange potential, with interesting consequences on the associated scalar spectrum. More recently, Enkhbat et al. gave a preliminary discussion of the fourth generation Yukawa bound states phenomenology at the LHC, considering a ultra-heavy degenerate new quark family \cite{Enkhbat:2011vp}. Although binding energies of Yukawa origin certainly provide a more consistent framework for a dedicated analysis, we restrict the present discussion to the context of Coulomb-like potential models, and assume that the short-distance behaviour of the quark-antiquark potential dominates (we refer the reader to \cite{Barger:1987xg} and \cite{Kuhn:1987ty} for previous and more exhaustive phenomenological reviews). In this context, the effects of the quarkonium wave function must be taken into account as we consider the corresponding production and decay modes. In perturbative QCD, the dominant contribution to $\eta _{Q}$ production cross-section in $pp$\ collisions reads \cite{Barger:1987xg}%
\begin{eqnarray}
\sigma (pp &\rightarrow &gg\rightarrow \eta _{Q})=\frac{\pi
^{2}\tau }{8m_{\eta }^{3}}\Gamma (\eta_{Q} \rightarrow gg)\text{ }\text{ }%
[\tau \frac{dL }{d\tau }]_{gg}, \\ \label{gluon}
\Gamma (\eta_{Q}  &\rightarrow &gg)=\frac{8\alpha _{s}^{2}(m_{\eta }^{2})}{%
3m_{\eta }^{2}}|R_{S}(0)|^{2},
\end{eqnarray}%
where $m_{\eta }=2m_{Q}$, and $[\tau \frac{dL }{d\tau }]_{gg}$ denotes the gluon luminosity with $%
\tau =m_{\eta}^{2}/s$. The full NLO partonic cross-sections for the $gg$, $qg$ and $q\bar{q}$ initiated reactions are provided in \cite{Kuhn:1992qw}, and have been updated in \cite{Arik:2002nd}. Interestingly, they are all proportional to the square of the $Q\bar{Q}$ radial wave function $R_{S}$ at the origin, which might drive the heavy quarkonium production cross-section $\sigma (gg\rightarrow \eta _{Q})\ $ to be substantially smaller than $\sigma(gg\rightarrow Q\bar{Q})$. Indeed, the quarkonium production rate typically amounts to a few percent of the pair production cross-section \cite{Kuhn:1992qw,Arik:2002nd,Hagiwara:2008df}. 

Depending whether the decay rate of the constituent particles, $\Gamma_{Q}$, is larger or smaller than the bound state annihilation rate, two different cases can be considered as the bound states can either undergo single quark decays or annihilate hadronically. As detailed in \cite{Kats:2009bv}, the strength of the annihilation decay signal is enhanced when the intrinsic width of the heavy quark $Q$ is decreased. Should its rate be of the same order of magnitude as the binding energy of the $\eta_{Q}$ state, the quarkonium can be broad and display little evidence of any resonance
behaviour over the continuum. On the other hand, if its 2-body decays are suppressed due to kinematical suppression or small couplings as discussed in Section \ref{sec:dec}, $Q$ can only decay through off-shell intermediate states. In such a case, the signature would correspond to an annihilation signal, namely, a hadronic final state markedly distinguishable from pair production and decay.

\begin{figure}[htbp]
\begin{center}
\includegraphics[width=0.50\textwidth]{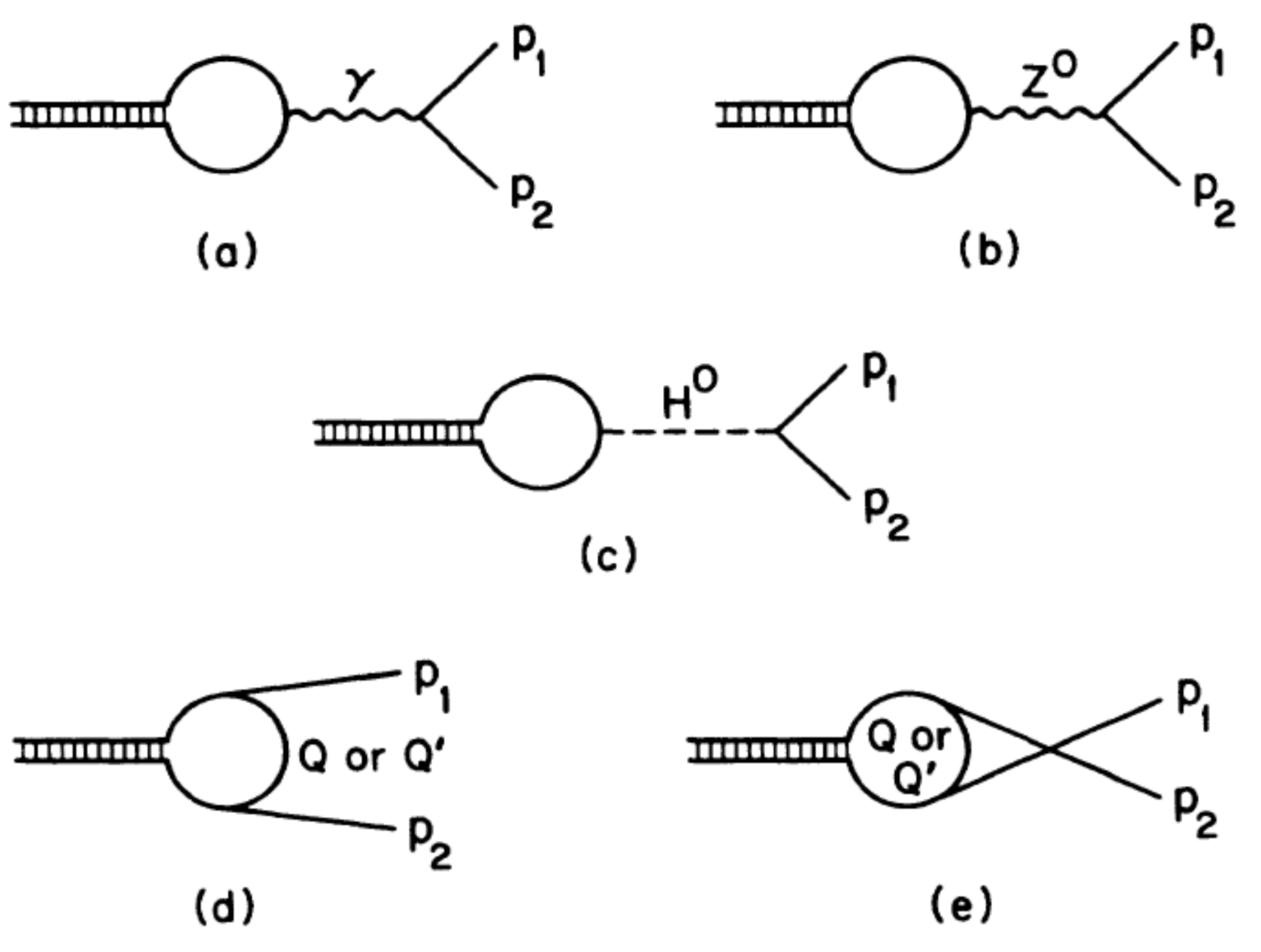}
\end{center}
\caption{First order diagrams for the $\eta_{Q}\rightarrow p_{1} p_{2}$ decays, where $p_{1}$ and $p_{2}$ can be a gauge boson pair, a fermion-antifermion pair, a gauge-Higgs boson pair, or a pair of Higgs bosons. The strong decay mode $\eta_{Q}\rightarrow gg$ is also allowed but not shown above. The (a), (b), and (c) diagrams denote decays via $\gamma$, $Z$ or $H$ exchange in the $s-$channel, while (d) and (e) correspond to the $p_{1}p_{2}$ decays involving quark exchange in the $t-$ and $u-$ channels. Figure from \cite{Barger:1987xg}.}
\label{Diagrams}
\end{figure}

We now give a brief and qualitative description of some of the expected $\eta _{Q}$ hadronic signatures at the LHC, considering either a chiral or vector-like heavy quark $Q$. While similar results hold for all $S$- and $P$-wave quarkonium states, we limit our analysis to the case of a stable, neutral $J^{PC}=0^{-+}$ pseudoscalar state $\eta _{Q}$. In particular, a $^{1}S_{0}$ quarkonium state formed from down-type fourth generation $b^{\prime}$ quarks is known to provide a possible candidate if the condition (\ref{Bigi}) is satisfied \cite{Arik:2002nd}. The latter is mainly produced through the gluon fusion process (\ref{gluon}) with a production cross-section two orders of magnitude larger than the $J^{PC}=1^{--}$ vector state \cite{Arik:2002nd}. If $b^{\prime}$ is the lightest fourth generation quark, a $\eta_{b^{\prime}}$ bound state can decay either through $q-b^{\prime}$ family mixing with $q=u,c,t$, or to boson pairs. Whether the single quark decays would compete with the $\eta_{b^{\prime}}$ hadronic modes depends on the heavy quark masses and mixings. If $m_{b^{\prime}} > 1$ TeV/$c^2$, a fourth-generation down-type quark dominantly decays into fermion-antifermion pairs if $|V_{qb^{\prime}}| \gtrsim 10^{-2}$. For $m_{b^{\prime}}<1$ TeV/$c^2$, the $\eta_{b^{\prime}}$ total width lies below $1$ GeV/$c^2$, and the quarkonium state decays proceed dominantly via the annihilation diagrams depicted in Figure \ref{Diagrams} \cite{Arik:2002nd}. Decays to gauge boson pairs, fermion-antifermion pairs, gauge-Higgs and Higgs boson pairs can occur through $\gamma, Z$ and $H$ exchange in the $s-$ channel, or proceed via quark mixing in the $t-$ and $u-$ channels, if allowed. Decays to Higgs boson pairs ($\eta_{Q} \rightarrow HH$) are forbidden by CP conservation in the case of $J^{PC}=0^{-+}$ pseudoscalars. While the processes involving charged currents are suppressed in the case of small quark couplings, the $\eta_{b^{\prime}}\rightarrow WW$ mode is allowed at loop-level via the exchange of the $SU(2)_{L}$ partner $t^{\prime}$, even if $|V_{tb^{\prime}}|\simeq 0$. Decays to $WW$ pairs are not allowed for $\eta_{T,B}$ bound states involving singlet vector-like quarks. 

\begin{figure}[t]
\begin{center}
\hspace{-0.5in}
\includegraphics[width=0.49\textwidth]{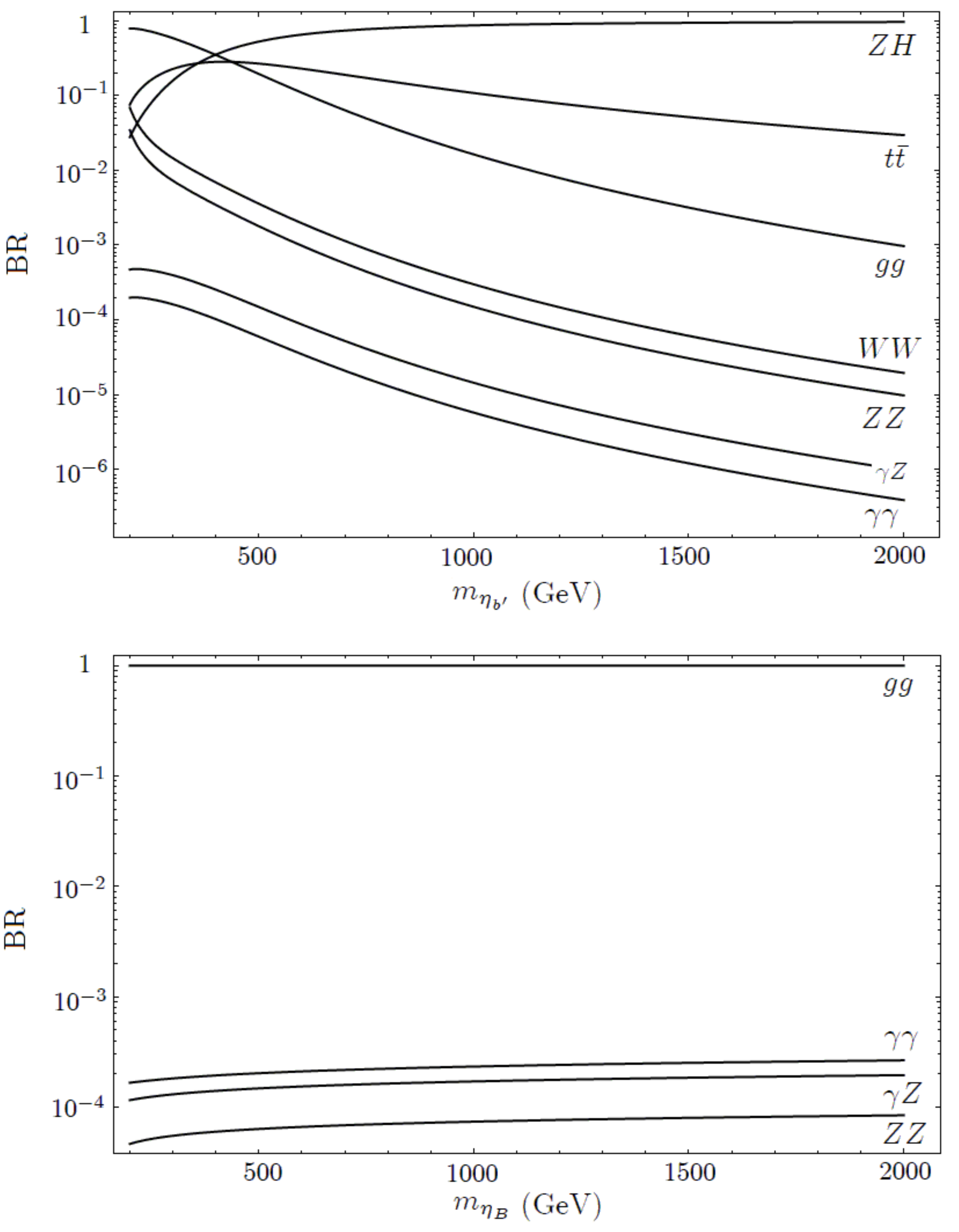}
\end{center}
\caption{Branching ratios as given in \cite{Kuhn:1993cp} for $\eta_{Q}$ quarkonium, considering a chiral $b^{\prime}$ quark (top) and a down-type singlet vector-like $B$ quark (bottom), with $m_{H}=125$ GeV/$c^2$.
$|V_{tb^{\prime}}|\simeq 1$ is assumed in the fourth generation case.}
\label{MirkesPlots}
\end{figure}

An exhaustive study of the production and subsequent decays of the $S-$ and $P-$ wave quarkonium states of a heavy chiral quark is available in \cite{Barger:1987xg} and \cite{Kuhn:1993cp}, where all the allowed production mechanisms and decay patterns have been thoroughly investigated. As an illustration, we compare in Figure \ref{MirkesPlots} the branching ratios of the allowed decay modes for quarkonium states formed by a chiral fourth generation quark $b^{\prime}$ and a vector-like isosinglet $B$ quark  \cite{Kuhn:1993cp}. The quarkonium annihilation decay rates depend markedly on the mass scale.

%\begin{figure}[htbp]

In the chiral case, the main decay modes are $ZH$, $gg$, $WW$, $ZZ$, $Z\gamma$ and $\gamma \gamma$ respectively. The fermionic decays $\eta _{b^{\prime}} \rightarrow q\bar{q}$ are also allowed if $|V_{qb^{\prime}}|\neq 0$. Substantial decay rates to $q \bar{q}$ fermion pairs are allowed in case of large $Q-q$ couplings. For pseudoscalar $\eta _{b^{\prime}}$ masses lighter than $450$ GeV/$c^2$, the dominant annihilation mode proceeds via the strong interaction, \emph{i.e.}, $\eta _{b^{\prime}}\rightarrow gg$. For larger masses, the width into two gluons is decreasing, whereas the annihilation mode into $ZH$ pairs takes over. This follows from the fact that the decay rates into Higgs and longitudinal gauge bosons are enhanced by the large Yukawa coupling of the heavy $b^{\prime}$ quark, so that the $ZH$ branching ratio becomes sizeable for increasing masses. The latter channel then leads to a salient signature for heavy fourth generation bound states. 

While fourth generation pseudoscalar quarkonia decays allow for striking signals, $T\bar{T}$ and $B\bar{B}$ bound states of vector-like quark singlets can only decay to $gg$, $\gamma \gamma$, $\gamma Z$ and $ZZ$ pairs, as the $\eta_{Q}\rightarrow ZH$ mode is absent \cite{Kuhn:1993cp}. This follows from the fact that the $ZH$ mode proceeds through the axial part of the neutral current coupling, which in turn is proportional to the third component of the weak isospin. Only $\gamma \gamma$, $Z\gamma$ and $ZZ$ would then signal the visible hadronic modes with branching ratios of about $10^{-4}$. $T\bar{T}$ and $B\bar{B}$ bound states of $SU(2)_{L}$ vector-like singlet quarks are thus hardly observable at the LHC as the gluon mode dominates. 
On the other hand, $T$ and $B$ vector-like doublet (triplet) partners with non-vanishing weak isospin can have large branching ratios to $ZH$ via $\gamma$, $Z$ or $H$ exchange in the $s-$channel. The $\eta_{X}$ and $\eta_{Y}$ quarkonia provide a noticeable exception, given that the exotic quarks $X_{5/3}$ and $Y_{-4/3}$ do not couple to neutral currents at tree-level. Since they only mix with the other states via charged currents, the annihilation modes including $\gamma$, $Z$ and $H$ exchange are thus forbidden for $\eta_{X}$ and $\eta_{Y}$, which then mainly decay to $gg$ and $WW$ bosons. Although it is suppressed below the percent level, the distinctive diphoton decay mode also provides a possible golden channel for discovery. If new stable, coloured particles are produced at the LHC, the resulting final states could indeed allow for resonant signatures including photons and leptons, from which their quantum numbers and masses can be identified independently of their decay modes. As the decay rate for such bound states might be small, model-independent searches for new resonances could be competitive with the direct searches \cite{Kats:2009bv}.

\subsection{Signatures for heavy stable particles : open-flavour mesons \label{sec:OPENsignatures}}

While $Q\bar{Q}$ quarkonium resonances might be challenging to observe, the possibility for \textquotedblleft open-flavour\textquotedblright hadrons $Q\bar{q}$ ($\bar{Q}q$) and $Qqq$ ($\bar{Q}\bar{q}\bar{q}$), with $q$ being a light SM quark, also cover a broad spectrum of new heavy mesons and baryons to search for. Should they be produced at the LHC, such states can form by capturing a light quark partner from the vacuum. As they pass through the detector, they can transform into various slowly-moving heavy states. Assuming that they hadronise with $u$, $d$ and $s$ quarks after being produced, Table \ref{tab:Qqbarmesons} lists the corresponding allowed stable mesons and baryons.
%which could be observed in colliders. 

The properties of new hypothetical fourth generation quarkonium and open-flavour mesons have been examined previously in \cite{Ikhdair:2005jj} considering large-$N$ expansion techniques. In \cite{Bashiry:2012zz}, their masses and decay constants have been estimated from the experimental measurements of the ordinary $b\bar{b}$ ($c\bar{c}$) hadron masses, using the QCD\ sum rules. Interestingly, both approaches indicate that the corresponding spectra share numerous features with most of the scenarios predicting stable bound states of heavy quarks beyond the Standard Model, depending on their masses, charges and angular momenta. 

Interestingly, the interactions of such open-flavour mesons with the material occur to be similar to that of R-hadrons, \emph{i.e.}, stable hadrons composed of a supersymmetric particle and at least one SM quark (see \cite{Kraan:2004tz,Mackeprang:2006gx} and references therein for related works). The 1/2 difference in spin with respect to stop hadrons has little impact on the search strategies, which depend mostly on the interactions between the lighter quarks and the matter of the detector. This is in substance the description given by the spectator model \cite{Bjorken:1977md,Suzuki:1977km}, for which the allowed transitions of heavy and light quarks within a given bound state are known to be flavour and spin independent \cite{Isgur:1989vq,Isgur:1991wq}. In this context, the heavy quark $Q$ can be considered as a source of kinetic energy which sole function is to give mass and momentum to the underlying hadron. Such new heavy bound states then behave as rather passive objects, consisting of a non-interacting heavy component, accompanied by lighter constituents ($u$ and $d$ quarks, mostly). While the partons are being scattered within the detector, their cross-sections vary with their inverse mass in perturbative QCD \cite{Fairbairn:2006gg,Mackeprang:2009ad}. A heavy $Q\bar{q}$ ($Q \bar{q}$) bound state is then expected to suffer small energy losses when interacting with the material, given that only its light quark half is responsible for the hadronic interactions with the detector.

We list in Table \ref{tab:ProcessList} the various allowed $2\rightarrow 2$ and $2 \rightarrow 3$ transitions involving meson-to-meson and meson-to-baryon conversions. In general, these interactions lead to simultaneous pion emission as the initial mesons convert into slowly-moving baryons.

\begin{table*}[htbp]
\begin{center}
\begin{tabular}{c|cc}
\hline\hline
Charge & Mesons & Baryons \\ \hline
&  &  \\ 
$Q=0$ & $\mathbf{T\bar{u}}, \mathbf{\bar{T}u}, \mathbf{B\bar{d}}, \mathbf{\bar{B}d}, B\bar{s}, \bar{B}s$ & $%
Tdd, Tds, Tss, \mathbf{Bud}, \bar{B}\bar{u}\bar{d}, Bus, \bar{B}\bar{u}\bar{s}, Yuu,\bar{Y}%
\bar{u}\bar{u}$ \\ 
$Q=1$ & $\mathbf{X\bar{u}}, \mathbf{T\bar{d}}, T\bar{s}, \mathbf{\bar{B}u}, \mathbf{\bar{Y}d}, \bar{Y}s$ & $%
Xdd, Xds, Xss, \mathbf{Tud}, Tus, Buu, \bar{B}\bar{d}\bar{d}, \bar{B}\bar{d}\bar{s}, \bar{B}%
\bar{s}\bar{s}, \bar{Y}\bar{u}\bar{d}, \bar{Y}\bar{u}\bar{s}$ \\ 
$Q=2$ & $\mathbf{X\bar{d}}, X\bar{s}, \mathbf{\bar{Y}u}$ & $\mathbf{Xud}, Xus, Tuu, \bar{Y}\bar{d}\bar{d},%
\bar{Y}\bar{d}\bar{s}, \bar{Y}\bar{s}\bar{s}$ \\ 
$Q=3$ & - & $Xuu$ \\ \hline\hline
\end{tabular}%
\caption{$Q \bar{q}$ ($\bar{Q}q$) mesons and $Qqq$ ($\bar{Q} \bar{q} \bar{q}$) (anti-)baryons involving $Q = X_{5/3}$, $T_{2/3}$, $B_{-1/3}$ and $Y_{-4/3}$ vector-like quarks. The states in bold font are hadrons which yields are expected to be substantial at the LHC, as predicted in \cite{Mackeprang:2009ad} for penetration depths between 0 and 3 meters. For simplicity,
only the neutral and positively charged hadrons are displayed. }
\label{tab:Qqbarmesons}
\end{center}
\end{table*}

\begin{table*}[htbp]
\begin{tabular}{c}
\hline\hline
Meson-to-Meson $2\rightarrow 2$ processes: \\ \hline
Charge exchange: \\ 
$Q\bar{u}+p\longleftrightarrow Q\bar{d}+n$ \\ 
\\ 
Elastic scattering: \\ 
$Q\bar{q}+p/n\rightarrow Q\bar{q}+p/n$ \\ \hline\hline
\end{tabular}%
\text{ \ \ \ \ \ \ \ \ \ \ \ \ \ }%
\begin{tabular}{c}
\hline\hline
Meson-to-Baryon $2\rightarrow 2$ processes: \\ \hline
Baryon exchange: \\ 
$Q\bar{u}+n\rightarrow Qud+\pi ^{-}$ \\ 
$Q\bar{u}+n\rightarrow Qdd+\pi ^{0}$ \\ 
$Q\bar{u}+p\rightarrow Quu+\pi ^{-}$ \\ 
$Q\bar{u}+p\rightarrow Qud+\pi ^{0}$ \\ 
\\ 
$Q\bar{d}+n\rightarrow Qud+\pi ^{0}$ \\ 
$Q\bar{d}+n\rightarrow Qdd+\pi ^{+}$ \\ 
$Q\bar{d}+p\rightarrow Quu+\pi ^{0}$ \\ 
$Q\bar{d}+p\rightarrow Qud+\pi ^{+}$ \\ \hline\hline
\end{tabular}%
%\bigskip
\\
\vspace{1cm}
\begin{tabular}{c}
\hline\hline
Meson-to-Meson $2\rightarrow 3$ processes: \\ \hline
Inelastic scattering: \\ 
$Q\bar{u}+n\rightarrow Q\bar{d}+n+\pi ^{-}$ \\ 
$Q\bar{u}+n\rightarrow Q\bar{u}+n+\pi ^{0}$ \\ 
$Q\bar{u}+n\rightarrow Q\bar{u}+p+\pi ^{-}$ \\ 
\\ 
$Q\bar{u}+p\rightarrow Q\bar{d}+p+\pi ^{-}$ \\ 
$Q\bar{u}+p\rightarrow Q\bar{u}+p+\pi ^{0}$ \\ 
$Q\bar{u}+p\rightarrow Q\bar{u}+n+\pi ^{+}$ \\ 
$Q\bar{u}+p\rightarrow Q\bar{d}+n+\pi ^{0}$ \\ 
\\ 
$Q\bar{d}+n\rightarrow Q\bar{u}+n+\pi ^{+}$ \\ 
$Q\bar{d}+n\rightarrow Q\bar{d}+n+\pi ^{0}$ \\ 
$Q\bar{d}+n\rightarrow Q\bar{u}+p+\pi ^{0}$ \\ 
$Q\bar{d}+n\rightarrow Q\bar{d}+p+\pi ^{-}$ \\ 
\\ 
$Q\bar{d}+p\rightarrow Q\bar{u}+p+\pi ^{+}$ \\ 
$Q\bar{d}+p\rightarrow Q\bar{d}+p+\pi ^{0}$ \\ 
$Q\bar{d}+p\rightarrow Q\bar{d}+n+\pi ^{+}$ \\ \hline\hline
\end{tabular}%
\text{ \ \ \ \ \ \ \ \ \ \ \ \ \ }%
\begin{tabular}{c}
\hline\hline
Meson-to-Baryon $2\rightarrow 3$ processes: \\ \hline
Baryon exchange: \\ 
$Q\bar{u}+n\rightarrow Qud+\pi ^{-}+\pi ^{0}$ \\ 
$Q\bar{u}+n\rightarrow Qdd+\pi ^{-}+\pi ^{+}$ \\ 
$Q\bar{u}+n\rightarrow Qdd+\pi ^{0}+\pi ^{0}$ \\ 
\\ 
$Q\bar{u}+p\rightarrow Quu+\pi ^{-}+\pi ^{0}$ \\ 
$Q\bar{u}+p\rightarrow Qud+\pi ^{-}+\pi ^{+}$ \\ 
$Q\bar{u}+p\rightarrow Qud+\pi ^{0}+\pi ^{0}$ \\ 
$Q\bar{u}+p\rightarrow Qdd+\pi ^{+}+\pi ^{0}$ \\ 
\\ 
$Q\bar{d}+n\rightarrow Qud+\pi ^{-}+\pi ^{+}$ \\ 
$Q\bar{d}+n\rightarrow Qud+\pi ^{0}+\pi ^{0}$ \\ 
$Q\bar{d}+n\rightarrow Qdd+\pi ^{0}+\pi ^{+}$ \\ 
\\ 
$Q\bar{d}+p\rightarrow Quu+\pi ^{-}+\pi ^{+}$ \\ 
$Q\bar{d}+p\rightarrow Quu+\pi ^{0}+\pi ^{0}$ \\ 
$Q\bar{d}+p\rightarrow Qud+\pi ^{0}+\pi ^{+}$ \\ \hline\hline
\end{tabular}%
%\bigskip
\caption{Meson-to-Meson and Meson-to-Baryon processes for $Q\bar{u}$ and $Q\bar{d}$ mesonic states, where $Q=X_{5/3},T_{2/3},B_{-1/3},Y_{-4/3}$ and $q=u,d$. Similar transitions can be obtained for the charge conjugated processes involving $\bar{Q} q$. Assuming that the cross-sections for all the transitions are of the same order of magnitude, Meson-to-Baryon processes to $Qud$ states are more likely than those involving $Quu$ and $Qdd$ in the final state.\label{tab:ProcessList}}
%\bigskip
\end{table*}

While they interact with the nuclear matter, most of the new states suffer multiple scattering and convert into baryons, allowing for $Quu$, $Qud$ and $Qdd$ states, as new heavy mesons are kinematically favoured to increase their baryon number by emitting one or more pions \cite{Mackeprang:2009ad}. Due to the lack of these in the material, the reverse reaction is known to be less favoured. 

Interestingly, new mesons which convert at the beginning of the scattering chain could generate tracks with \textquotedblleft dashed-lines\textquotedblright, signalling possible baryonic or electric charge exchange. Indeed, $Q\bar{q}$ and $\bar{Q}q$ bound states traversing a medium composed of light quarks likely flip their electric charge, frequently interchanging their parton constituents with those of the material nuclei. Eventually, they could leave the detector as heavy stable neutral particles and hence not be observable in muon detectors. The modeling of the nuclear interactions of new heavy hadrons traveling through matter, a survey of which is given in \cite{Mackeprang:2009ad,deBoer:2007ii}, actually favours scenarios with significant charge suppression if heavy open-flavour mesons have a sizeable probability to transform into neutral particles while traversing the detector. This poses a serious challenge for the experimental searches, since the most conservative scenarios assume a complete charge suppression where 100\% of the produced hadrons become neutral before reaching the muon detectors. Taking a specific example, we consider pair-produced $T_{2/3}$ quarks hadronising into the heavy states $T \bar{q}$ and $\bar{T} q$ immediately after production. For convenience, we assume that a majority ($\gtrsim 90\%$) of the new heavy quarks initially hadronise into mesons, while a smaller amount ($\lesssim 10\%$) form baryons \cite{Fairbairn:2006gg}. 
As they interact farther in the detector, most of the $T\bar{q}$ mesons eventually exit the detector as $T\bar{u}$ and $T\bar{d}$ mesonic states, or as $Tud$ baryons. Baryon-to-meson conversions are allowed as well, as these states can also annihilate back into $T\bar{u}$ and $T\bar{d}$ mesons, yet with a smaller rate. $\bar{T}q$ mesons of vector-like antiquarks, on the other hand, unlikely give rise to antibaryons, but can still flip their charge through the exchange of $u$ and $d$ quarks with the material. Possibly large flavour fractions for $\bar{T}u$ and $\bar{T}d$ can thus be expected in the detector, with a negligible amount of $\bar{T} \bar{q} \bar{q}$ states. If such antibaryons were to form, 
they would quickly annihilate by baryon-to-meson interactions, producing pions \cite{Kang:2007ib}.

After travelling through the detectors a few nuclear lengths away from the production vertex, the novel hadrons transform in roughly 100\% to the positively charged baryon $Tud$, whereas one half of the $\bar{T}-$ states thus transform into $\bar{T}u$ mesons, and the other half in $\bar{T}d$. The corresponding hadronic flavour decompositions, evaluated as a function of the penetration depth in the detector, can be read from fig. 5 of \cite{Mackeprang:2009ad} for R-hadrons involving stable scalar top and antitop quarks. We notice that similar results are obtained when replacing stop $\tilde{t}$ squarks by stable quarks $T$ of equal mass. The $Tud$, $T\bar{u}$ ($\bar{T}u$) and $T\bar{d}$ ($\bar{T}d$) states are then expected to retain the largest flavour composition fractions when considering penetration depths larger than $3$ meters, and are thus the most likely observable states, which we emphasised in bold font in Table \ref{tab:Qqbarmesons}.

Except for $Yuu$ and $\bar{Y}\bar{u}\bar{u}$, it is interesting to notice that none of the states listed in Table \ref{tab:Qqbarmesons} is neutral if involving vector-like quarks with exotic charges. The corresponding bound states might give rise to possibly large fractions of slowly moving charged particles $Xud$ and $Yud$ with $Q=+2e$ and $-e$ respectively, accompanied by a small amount of $X\bar{u}$, $X\bar{d}$, $Y\bar{u}$ and $Y\bar{d}$ mesons (and charge conjugates). These particles are all electrically charged as they go across the detector, emitting charged and neutral pions in the process together with losing small amounts of energy. Given the small expected interaction cross-section from the Meson-to-Baryon transitions, the processes shown in Table VI leave small energy deposits in the calorimeters, of the order of $O(1)$ GeV \cite{Fairbairn:2006gg}. On the other hand, such stable hadrons would lead to observable tracks due to the ionisation energy losses, allowing for signatures similar to slow-moving muons with high transverse momentum. 
Searches for stable charged particles, if adapted to such a case, thus provide a promising strategy to rule out the possibility for novel exotic quarks with large lifetimes. Given these very specific signals, the aforementioned signatures can certainly be discriminated at the LHC.

\section{Experimental aspects \label{sec:experiment}}

In this section, we review the past and current experimental searches for new long-lived quarks. Limitations of these searches are then discussed along with possible extensions to enhance sensitivity to heavy quarks. Reinterpretations of direct searches for heavy quarks as well as general searches for long lived particles are  given in the context of long lived heavy quarks.

\subsection{Previous searches at Tevatron \label{sec:Tevatron}}

The searches undertaken at the Tevatron resulted in already stringent mass bounds on long-lived heavy quarks, yet not without assumptions. Looking for long-lived parents of the $Z$ boson in displaced vertices from $p\bar{p}$ collisions at $\sqrt{s}=1.8$ TeV, CDF set limits on the cross-section of a fourth generation charge $-1/3e$ quark as a function of its lifetime with an integrated luminosity of $90$ pb$^{-1}$ \cite{Abe:1998ee}.\ Isolated electron-positron pairs originating from $Z$ decays were searched for in the exclusive channel $b^{\prime }\rightarrow bZ$ with $m_{b^{\prime }}>m_{Z}$. Finding no evidence for new long-lived particles, CDF excluded $m_{b^{\prime }}<148$ GeV/$c^{2}$ for $c\tau =$ 1 cm at 95\%\ CL. This limit drops to $m_{Z}+m_{b}\simeq 96$ GeV/$c^2$ if $c\tau >$ 22 cm or $c\tau <$ 0.009 cm. Previously, the D0 collaboration already ruled out the range $m_{Z}/2<m_{b^{\prime}}<m_{Z}+m_{b}$ from $b^{\prime }\rightarrow b\gamma $ searches in \cite{Abachi:1996fs} for all proper lifetimes, while a $b^{\prime }$ quark with a mass lower than $m_{Z}/2$ was previously dismissed by the LEP direct searches \cite{Decamp:1989fm}.
Within a 90\ pb$^{-1}$ data sample of $p\bar{p}$ collisions recorded during 1994-95, CDF performed a search for low velocity massive charged stable particles leaving large amounts of energy in the calorimeters \cite{Acosta:2002ju}. 

Assuming a muonlike penetration and searching for an anomalously high ionisation energy loss signature, the data was found to agree with background expectations, and upper limits of the order of 1 pb were derived on the production cross-section. Sensitive to long-lived fourth generation quarks scenarios, the lower bounds $m_{b^{\prime }}>190$ GeV/$c^{2}$ and $m_{t^{\prime }}>220$ GeV/$c^{2}$ have been obtained for $q=-1/3e$ and $q=2/3e$ stable quarks respectively, with no observed excess over background. 
More recently, D0\ studied a 1.1 fb$^{-1}$ data sample and looked for $Z\rightarrow e^{-}e^{+}$ decays assumed to follow from a long-lived $b^{\prime }$ quark parent with BR($b^{\prime }\rightarrow bZ)=100\%$ \cite{Abazov:2008zm}. Assuming that the electromagnetic showers were originating from the same vertex, the analysis found no hint away from the $p\bar{p}$ interaction point. D0 excluded $m_{b^{\prime}}<190$ GeV/$c^2$ at the 95\% confidence level for decay lengths between 3.2 mm and 7 m.
With the same integrated luminosity, CDF\ excluded $m_{b^{\prime }}<268$ GeV/$c^2$ at 95\% C.L., considering a long-lived $b^{\prime }$ quark decaying exclusively into a $Z$ boson and a $b$ jet \cite{Aaltonen:2007je}. 
However, it has been emphasised in \cite{Hung:2007ak} that the assumption BR($b^{\prime }\rightarrow bZ)=100\%$ is inaccurate for $m_{b^{\prime }}>255\ $GeV/$c^2$, given that the $b^{\prime }$ decay mode $b^{\prime }\rightarrow tW$ should take over if $m_{b^{\prime }}>m_{t}+m_{W}$. Furthermore, the processes $b^{\prime }\rightarrow (u,c)W$ were hinted to proceed with non-negligible rates, leading to an even lower branching ratio. Indeed, even if the $b^{\prime }\rightarrow tW$ decay dominates, the aforementioned mass limits depend sensitively on the CKM mixing elements between the fourth and the first three generations. If non-unity couplings between the fourth and the lighter quarks are allowed, the CDF lower bound on $m_{b^{\prime }}$ can be significantly affected.

Additionally, it is known that the above CDF limits do not apply for long-lived heavy quarks decaying between roughly 1 cm and 3 m within the detector \cite{Hung:2007ak}. Should their couplings to SM quarks lie in the corresponding range $4.5\times 10^{-9}-7.8\times10^{-8}$, there exists no bounds on the $t^{\prime }$ and $b^{\prime }$ masses in this uncovered region. For decay lengths larger than the detector dimensions, the lower bounds drop to $m_{t^{\prime }}>220$ GeV/$c^2$ and $m_{b^{\prime }}>190$ GeV/$c^2$, as obtained from the stable quark searches \cite{Acosta:2002ju,Abazov:2008zm}. 
%Although most experiments require the quarks to decay within a few centimeters from the beam pipe, this is generally not the case for new heavy particles having very small mixings with the SM quarks. This issue should be treated carefully at the LHC.
\subsection{LHC \label{sec:LHC}}

\subsubsection{Limitations of direct searches \label{sec:ExpDirect}}

While dedicated results for long-lived bound states of vector-like heavy quarks are still missing at the LHC, various searches for  heavy quarks of zero lifetime  have already been performed by the ATLAS and CMS experiments. In this section, we discuss how the results of searches for prompt production could be reinterpreted for  lifetimes larger than $10^{-10}$~s.

%Although their results usually depend on the assumption of exclusive branching ratios and ignore the possibility of relevant lifetimes, we will see that some of them can be applied to the case of very long-lived states leading to HSCP signatures.

The main aspects of  reinterpreting these results are the branching ratios to the investigated final states, which may be different in alternative models, and of course the lifetime of the heavy particle.

Considering here a specific example, the current best limit on $b^{\prime}$ production, published by the CMS collaboration, excludes production cross-sections of $\sigma > 0.1$~pb at 95\% C.L. with an integrated luminosity of 4.9 fb$^{-1}$ at $\sqrt{s} = 7$ TeV \cite{Chatrchyan:1438749}, assuming a branching ratio ${\rm BR}(b'\to tW) = 100\%$.  As we have seen in Section \ref{sec:dec}, the assumption of exclusive branching ratios is not generally valid for new heavy quarks. Nevertheless, the fraction to which a hypothetical signature occurs  can be used to recalculate the cross-section limit in the most na\"{\i}ve approximation
\begin{equation}
\sigma_{\rm true} = \sigma_{\rm excluded} \times \frac{ {\rm BR}_{\rm assumed}}{ {\rm BR}_{\rm true}},
\end{equation}
where $\sigma_{\rm excluded}$ is the excluded cross-section under the assumption of ${\rm BR}_{\rm assumed}$, and ${\rm BR}_{\rm true}$ is the branching ratio in the given model. 

While this statement is a trivial estimation, the reinterpretation for longer lifetimes requires more thought. In particular, the implicit assumption of heavy quarks decaying promptly at the production vertex is valid only for very short lifetimes. If these particles form bound states and propagate certain distances before their decay, they can potentially escape the direct searches. In such a scenario, the published limits could be considerably weaker, depending on the particle's lifetimes, simply because they would fly too far to be detected at the primary vertex. 
%An illustration will be given in Section \ref{sec:LongLIvedSearchesReview}.

For intermediate lifetimes with displaced decay vertices within the detector volume (cfr. the region (\emph{ii}) discussed in Section \ref{sec:sigDisplacedVert}), a recalculation of the published limits can  be attempted. Assuming an exponential decay function, a fraction of the decays will always happen in the vicinity of the beam line, so that the prompt searches will pick them up. Based on this fraction, one can recalculate the limits as a function of the heavy quark lifetime. However, in order to do this correctly, one requires the exact selection efficiency as a function of the displacement from the beam line. Unfortunately, such information has not been made available by the experiments. Still, it can be estimated from an educated guess under specific assumptions. For instance, the analysis in \cite{Chatrchyan:1438749} applies $b-$jet tagging, which requires high-quality tracks to originate within the inner detector volume. The CMS innermost part, the pixel detector consists of three barrel layers with the innermost layer at a distance of $4.4$ cm from the beam line. This represents the technical limitation to this analysis, so that we can assume that the selection efficiency vanishes at a transverse displacement of around $4$ cm.

Even in analyses without application of $b-$jet tagging, stringent quality cuts are usually required to select reconstructed detector objects originating from the primary vertex. One of the motivations for these requirements is the rejection of pileup. Jets from displaced decays which do not point to the production vertex are mostly  removed by the pileup cleaning procedures. We therefore make the   assumption that events are kept in the direct searches if the decay happens at less than $4$ cm from the primary interaction vertex, and lost otherwise. The fraction of the lost events can be determined from the distance these particles travel before their decay. This distance $d$ depends on the mass, the momentum and the proper lifetime $\tau$. It is defined in the laboratory frame as $d= \gamma \cdot \beta \cdot c \cdot \tau$. The fraction is obtained by convoluting $d$ with an exponential decay function and the momentum spectrum which determines $\gamma$ and $\beta$.
 
\begin{figure}[htbp]
\begin{center}
\includegraphics[width=0.46\textwidth]{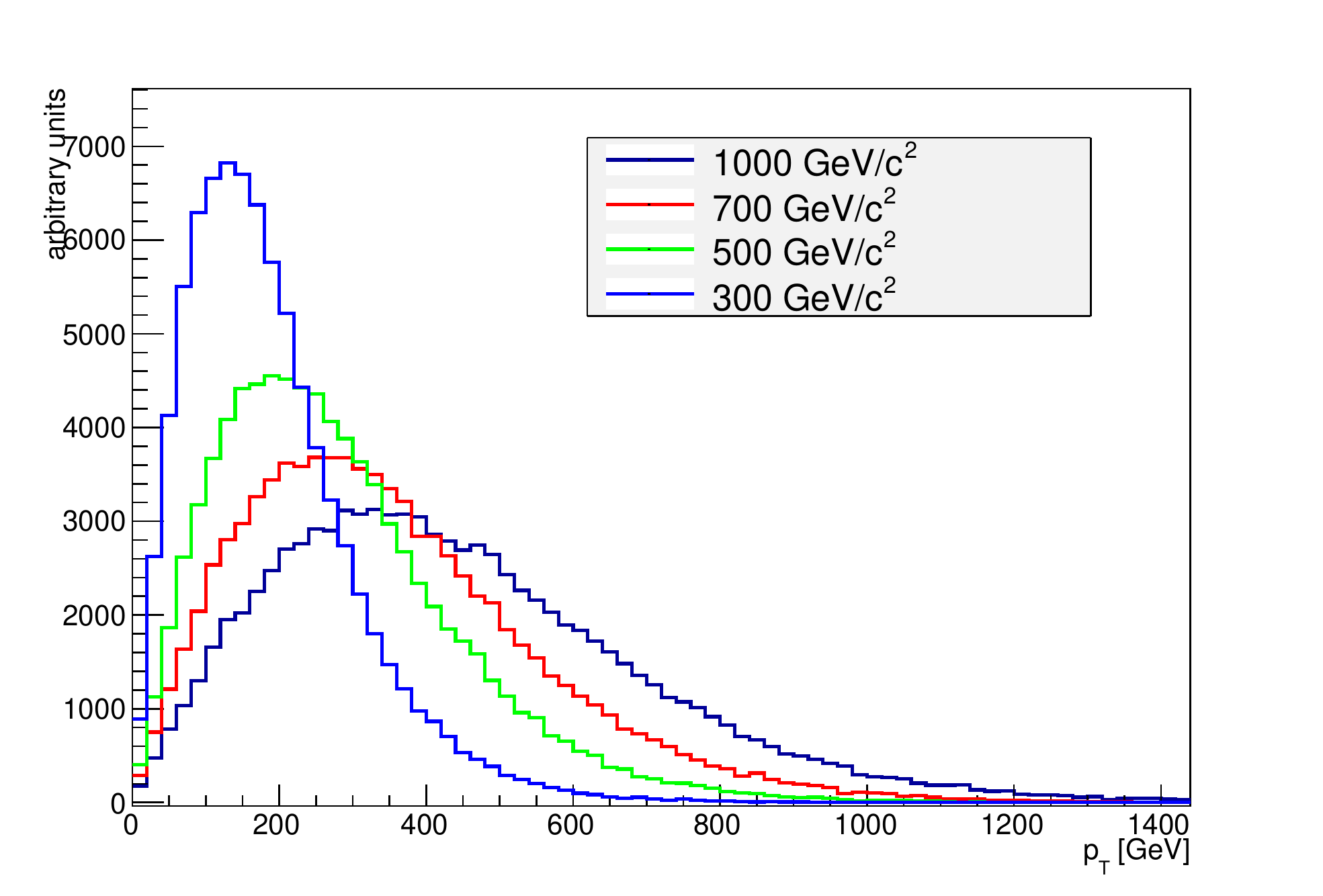}
\end{center}
\caption{Transverse momentum distribution of pair-produced heavy quarks as predicted by MadGraph.}
\label{fig:TprimeMomentum}
\end{figure}

\begin{figure}[htbp]
\begin{center}
\includegraphics[width=0.46\textwidth]{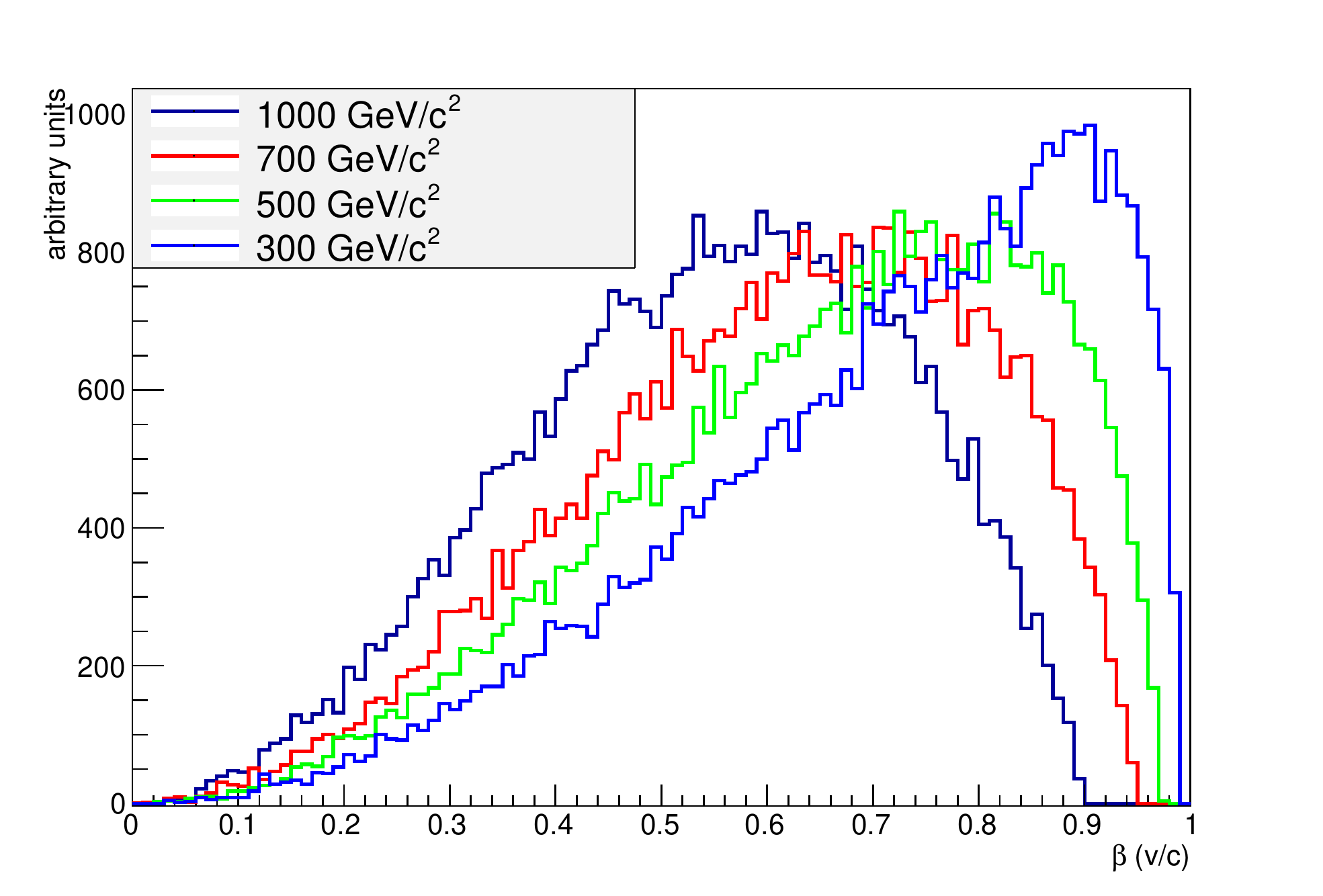}
\end{center}
\caption{Velocity distribution of pair-produced heavy quarks in units of $\beta = v/c$ as predicted by MadGraph.}
\label{fig:velocity}
\end{figure}

We take the momentum spectrum of pair-produced heavy quarks as predicted by the MadGraph \cite{Alwall:2011uj} Monte Carlo generator. The assumed center of mass energy is $\sqrt{s}=8$ TeV. The momentum is needed for a precise calculation of the decay distance on an event by event basis because the velocity may be significantly smaller than $c$.  The $p_T$ distribution for various quark masses is shown in Figure \ref{fig:TprimeMomentum}. 
 The distribution of the velocity in units of $c$ is shown in Figure \ref{fig:velocity}. 

\begin{figure}[htbp]
\begin{center}
\includegraphics[width=0.48\textwidth]{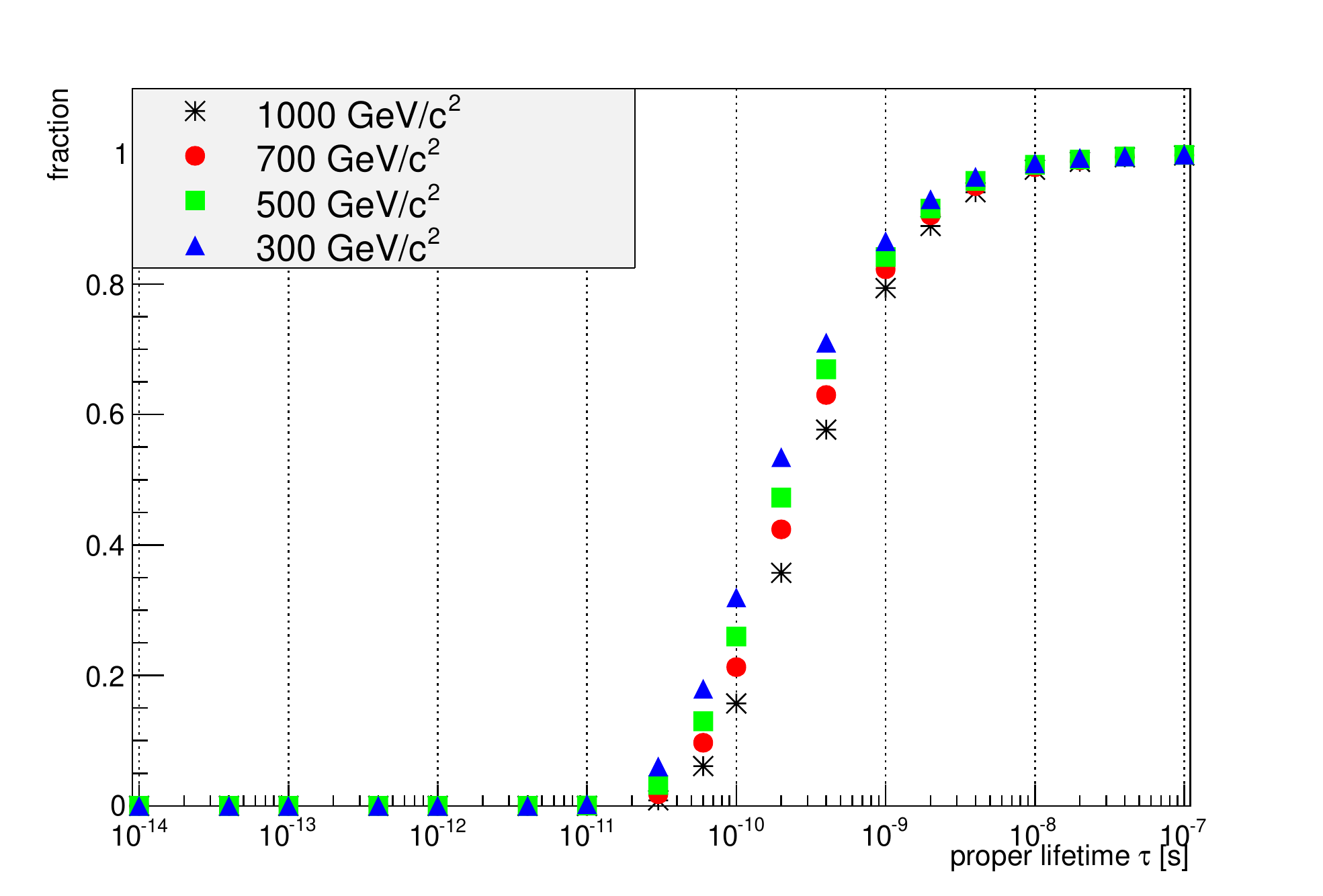}
\end{center}
\caption{Fraction of rejected decays of heavy quarks outside of the geometrical acceptance of $4$ cm around the
primary interaction vertex as a function of the proper lifetime $\tau$.}
\label{fig:TprimeSelEff4cm}
\end{figure}

Our simulation assumes that the production mechanism and kinematics are independent of the lifetime of the particle. Possible limitations of that assumption are discussed in Section \ref{sec:ExpDiscussion}. Calculating the distribution of decay vertices in the detector frame, we obtain the fraction of decays outside of the geometrical acceptance of $4$ cm as a function of the proper lifetime $\tau$. Table \ref{tab:TprimeSelEff4cmTable} summarises our results for four benchmark points in the parameter space. The results are displayed as a function of the proper lifetime in Figure \ref{fig:TprimeSelEff4cm}.
\begin{table}[htpb]
\begin{tabular}{lccc}
\hline\hline
Lifetime & $10^{-10}$ s &  $10^{-9}$ s & $10^{-8}$ s  \\ \hline
$m=300$ GeV/$c^{2}$ & 32\% & 86.5\% & 98.5\% \\ 
$m=500$ GeV/$c^{2}$ & 26\% & 84.1\% & 98.3\% \\ 
$m=700$ GeV/$c^{2}$ & 21\% & 82.3\% & 97.9\% \\ 
$m=1000$ GeV/$c^{2}$ & 16\% & 79.4\% & 97.5\% \\ \hline \hline
\end{tabular}
\caption{Fraction of rejected decays of heavy quarks outside of the geometrical acceptance of $4$ cm around the primary interaction vertex for three different lifetimes and four different heavy quark masses.}
\label{tab:TprimeSelEff4cmTable}
\end{table}

We conclude that direct searches are only valid for lifetimes which are considerably shorter than about $10^{-10}$ s, otherwise the particles would propagate too far and be rejected by the selection criteria. We also stress that this result is relatively independent of the particle masses within our approximations. 

The published limits on the heavy quark production cross-sections can  be simply reinterpreted in the region $10^{-8}$  s $< \tau < 10^{-10} $~s  by
\begin{equation}
 \sigma_{\rm true} = \sigma_{\rm excluded} \times (1-f)
\end{equation}
where  $f$ is the fraction of lost events given in Table \ref{tab:TprimeSelEff4cmTable} and Figure \ref{fig:TprimeSelEff4cm}.  The recalculation of the mass limits can then be done by comparing the recalculated cross-sections with predicted cross-sections as a function of the mass.

\subsubsection{Displaced vertices \label{sec:displaced}}
The immediate question that arises from the previous results is whether lifetimes larger than $10^{-10}$ s are covered by dedicated searches for long-lived particles, either through displaced vertices or signatures of stable particles propagating through the full detector.

Displaced topologies have been searched for at LHC and published, for instance in \cite{CMS-PAS-EXO-11-101} and \cite{Aad2012478}. These searches are indirectly applicable to heavy quarks  and will be discussed in this section. 
%We also suggest small modifications to these searches to extend their reach to long-lived heavy quarks.

The CMS analysis \cite{CMS-PAS-EXO-11-101} is looking for massive long-lived spinless neutral $\chi$ bosons, produced in decays of Higgs bosons ($H\to \chi\chi$). The $\chi$ bosons are decaying to di-leptons $\chi \to l^+l^-$. This analysis is feasible thanks to the CMS track reconstruction algorithm which is able to identify very displaced tracks not originating from the primary interaction vertex. By fitting tracks of two oppositely charged leptons  to a common vertex,  transverse displacements up to $50$ cm from the beam line can be reconstructed. This analysis is very sensitive and it is able to put limits on production cross-sections of the order of $\sigma \times \mathrm{BR < 10^{-3}}$ pb at 95\% C.L., depending on the assumed masses of the  $\chi$ bosons and their mother particles.
The decays of heavy quarks such as $Q\rightarrow tW$, for instance, can lead to  displaced di-lepton vertices as well, because both $W$ bosons can decay leptonically. 

In this analysis, the momentum of the vertex is required to be parallel to the vector pointing from the primary vertex to the $\chi$ boson decay vertex in the transverse plane. This ``collinearity'' cut is 0.2 (0.8) radians for muon (electron) final states.  A direct interpretation  of the published limits in context of long-lived heavy quarks is therefore not straightforward.  Due to the presence of the neutrinos and of the $b$-quark jet in $Q\rightarrow tW$ decays, the momentum of the di-lepton vertex will not be parallel to the vector pointing from the primary to the secondary vertex. However, we can attempt to estimate the efficiency of this collinearity cut based on simulated events. We use again our MadGraph simulation from Section \ref{sec:ExpDirect} and calculate the angle between the momentum  of a heavy quark $Q$ and the di-muon momentum in leptonic $Q\rightarrow tW$ decays.  It is found that the distribution of this collinearity angle is fairly independent  of the mass of the heavy quark. The collinearity cut has an efficiency of 68\% for electrons and 25\% for muons in leptonic $Q\rightarrow tW$ decays which represents a moderate decrease. A reinterpretation of the published results, taking these efficiencies into account, will be considered in the following.

A central component that would be necessary to derive trustworthy limits on heavy quark production from displaced vertex searches is the Monte Carlo simulation of the signal process. The interactions of these heavy particles  with the detector material and the efficiency of their reconstruction and selection need to be estimated in order to facilitate the calculation of limits. Difficulties may arise from missing implementations of certain exotic models in the simulation software. Existing simulations of heavy stable particles such as those of R-hadrons may provide good approximations though \cite{Mackeprang:2009ad}.

 We see in \cite{CMS-PAS-EXO-11-101} that different assumptions for the masses $m_H$ and $m_\chi$ lead to different exclusion limits due to the kinematical properties of the final state. A direct matching of the mass of a heavy quark  to $m_H$ and $m_\chi$ is not straightforward. The assumption that the results from \cite{CMS-PAS-EXO-11-101} are applicable for heavy quarks can therefore only be understood as an approximation. In the following we assume that the limits for higher $m_H$ and $m_\chi$ values are more applicable to heavy quarks, so we only consider those mass points with $m_H \geq 400$ GeV/$c^2$ and $m_\chi \geq 50$ GeV/$c^2$. Fortunately,  the resulting exclusion limits vary only in a small window between $5\cdot 10^{-4}$ pb and $2\cdot 10^{-3}$ pb  in the dimuon channel for $1$ cm $ < \rm c\tau < 10$ cm when scanning the various mass points for $m_H$ and $m_\chi$. Following the strategy of a worst case scenario we use the worst limit of  $2\cdot 10^{-3}$ pb which weakens to $8\cdot 10^{-3}$ pb taking the effect of the collinearity cut into account. Now the $tW$ final state has only a 1\% branching ratio into dimuons which results in an excluded cross-section of $\sigma \cdot BR(Q\to tW) > 0.8$~pb for $1 $ cm $< \rm c\tau < 10$ cm. In the context of QCD pair production of heavy quarks, the value of 0.8~pb corresponds to a mass of $m_Q = 435$~GeV/$c^2$ at $\sqrt{s} = 7$~TeV \cite{Cacciari:2011hy}. This can therefore be assumed to be the exclusion limit under the assumption of a branching fraction $BR(Q\to tW) = 100\%$. Comparing this to  the searches at CDF and D0 quoted in Section \ref{sec:Tevatron} we see  a clear improvement of the limits, at least in the range $1 $ cm $< \rm c\tau < 10$ cm. While the reinterpretation is difficult due to the discussed reasons, much stronger bounds would certainly be possible by including the heavy quark signatures into the displaced vertex analysis.

A displaced vertex analysis has also been performed by the ATLAS collaboration \cite{Aad2013280},  following a slightly different strategy. Displaced vertices are reconstructed in an inclusive way, using all displaced tracks in the event as potential seeds. In contrast to CMS, where only opposite-sign di-muon vertices have been used, the ATLAS approach requires at least five tracks at the vertex and an invariant mass of the vertex of at least 10 GeV/$c^2$.  To ensure a good fit quality, only vertices within the fiducial barrel pixel detector volume are considered, which means that transverse displacements up to $18$ cm are considered. In addition, one muon with $p_T>50$ GeV/$c$ is required to be associated with the signal vertex.  This analysis has been carried out in the context of an $R$-parity violating supersymmetric scenario, deriving limits on $\sigma \cdot \mathrm{BR}^2$, where $\mathrm{BR}^2$  is the square
of the branching ratio for produced squark pairs to decay via long-lived neutralinos to muons and quarks.  The requirement that the triggering muon candidate is associated with the displaced vertex ensures that the selection efficiency for each neutralino is independent of the rest of the event. This facilitates a reinterpretation for scenarios with different numbers of long-lived neutralinos in the event. 

Also in this analysis the limits depend on masses and kinematics of the hypothetical long-lived particles and their mother particles. In the most pessimistic scenario  the excluded cross-sections are approximately  $\sigma \cdot \mathrm{BR}^2 > 10^{-2}$~pb for lifetimes of  3~mm $< {\rm c}\tau < 10^{2}$~mm  using an integrated luminosity of 4.4~fb$^{-1}$ at $\sqrt{s}=7$ TeV.  For even larger lifetimes of  $10^{2}$~mm $< {\rm c}\tau < 10^{3}$~mm the limits are better than $\sigma \cdot \mathrm{BR}^2 > 10^{-1}$~pb.

The interpretation of these results for decays such as the $Q\rightarrow tW$ or even $Q\rightarrow bW$ is possible because these signatures would certainly give rise to decay vertices with very high invariant mass and high track multiplicity. The $tW$ final states has at least one muon in 19\% of the cases which means that the excluded values are $\sigma \cdot BR(Q\to tW) > 0.05$~pb for 3~mm $< {\rm c}\tau < 10^{2}$~mm. This is about an order of magnitude better than our reinterpretation of the CMS result. The value of 0.05~pb corresponds to $m_Q = 650$~GeV/$c^2$ \cite{PhysRevD.81.035006}.

\subsubsection{Heavy Stable Charged Particles \label{sec:LongLIvedSearchesReview}}
For very long-lived scenarios (cfr. the region (\emph{iii}) discussed in Section \ref{sec:sigDisplacedVert}), the new heavy
states do not decay, but travel through the full detector. In the
following we review recent results by CMS \cite{Chatrchyan:1445273} along with an assessment of
their relevance for HSCPs. 

The main
identification criteria for HSCPs are high momentum, high ionization and
long time of flight (TOF).  The energy loss $dE/dx$ along the track is calculated from the charge deposits in the silicon tracker, while the TOF is obtained from the arrival time in the muon system. These quantities are uncorrelated for SM particles. This noncorrelation is used to estimate the background yields in signal-depleted regions. 

Models of charge suppression due to interaction of HSCPs with the detector material are considered as well. If charged particles become neutral while propagating through the detector they do not reach the muon chambers. The TOF criterion cannot be used in this case. Therefore, the results have also been estimated without the TOF requirement at the cost of weaker exclusion limits.

The obtained limits for a given mass can be very different depending on the assumed particle type. This is due to the predicted kinematics of the particle and its expected detector signal. For instance, the limit on pair production of a scalar top quark with a mass between $300$ and $800$ GeV/$c^2$ is $\sigma < $~3 fb for an integrated luminosity of $5$ fb$^{-1}$ at $\sqrt{s} = 7$ TeV.

Limits for long-lived heavy quarks have not been considered explicitly, unfortunately. However, as an approximation,  we make the assumption that a long-lived bound state of a stop would behave similar or equal to a bound state of a  heavy up-type quark \cite{Mackeprang:2009ad}. This concerns both the interaction with the detector material and the production mechanism which is assumed to proceed via the strong interaction. It is then possible to use  the stop exclusion limits directly for stable heavy quarks. Limitations in that assumption are discussed in Section \ref{sec:ExpDiscussion}.

The predicted production cross-sections of heavy quarks are quite large for low masses (Section \ref{sec:prod}). Even with a large systematic safety margin, these cross-sections can be considered to be excluded by the HSCP searches. The heavy quark pair production cross section for $m_Q=800$~GeV/$c^2$ (the highest mass value considered in \cite{Chatrchyan:1445273}), computed at NLO,  is $\sigma = 9.7$~fb at $\sqrt{s} = 7$ TeV \cite{Cacciari:2011hy}, which can therefore be assumed to be excluded for sufficiently long lifetimes. Heavy quark masses of the order of $m_{Q} \approx 1$ TeV/$c^2$ are not yet excluded though, neither in direct searches for prompt decays nor in long-lived searches.

To make a quantitative statement about the excluded lifetimes by the HSCP searches, we repeat our simulation from Section \ref{sec:ExpDirect}. To be detected by these analyses, the particles have to be produced at the primary vertex and they have to traverse the full tracking devices, and the muon chambers for the combined tracker+TOF analysis (or the tracker  for the tracker-only analysis). The CMS tracker has a length of about $6$ m and a diameter of $2.8$ m \cite{CMSpaper}. We can calculate the fraction of decays inside the tracker volume and assume that these decays will not be reconstructed in the HSCP analyses because of missing hits in the outermost tracker layers. 

\begin{figure}[htbp]
\begin{center}
\includegraphics[width=0.52\textwidth]{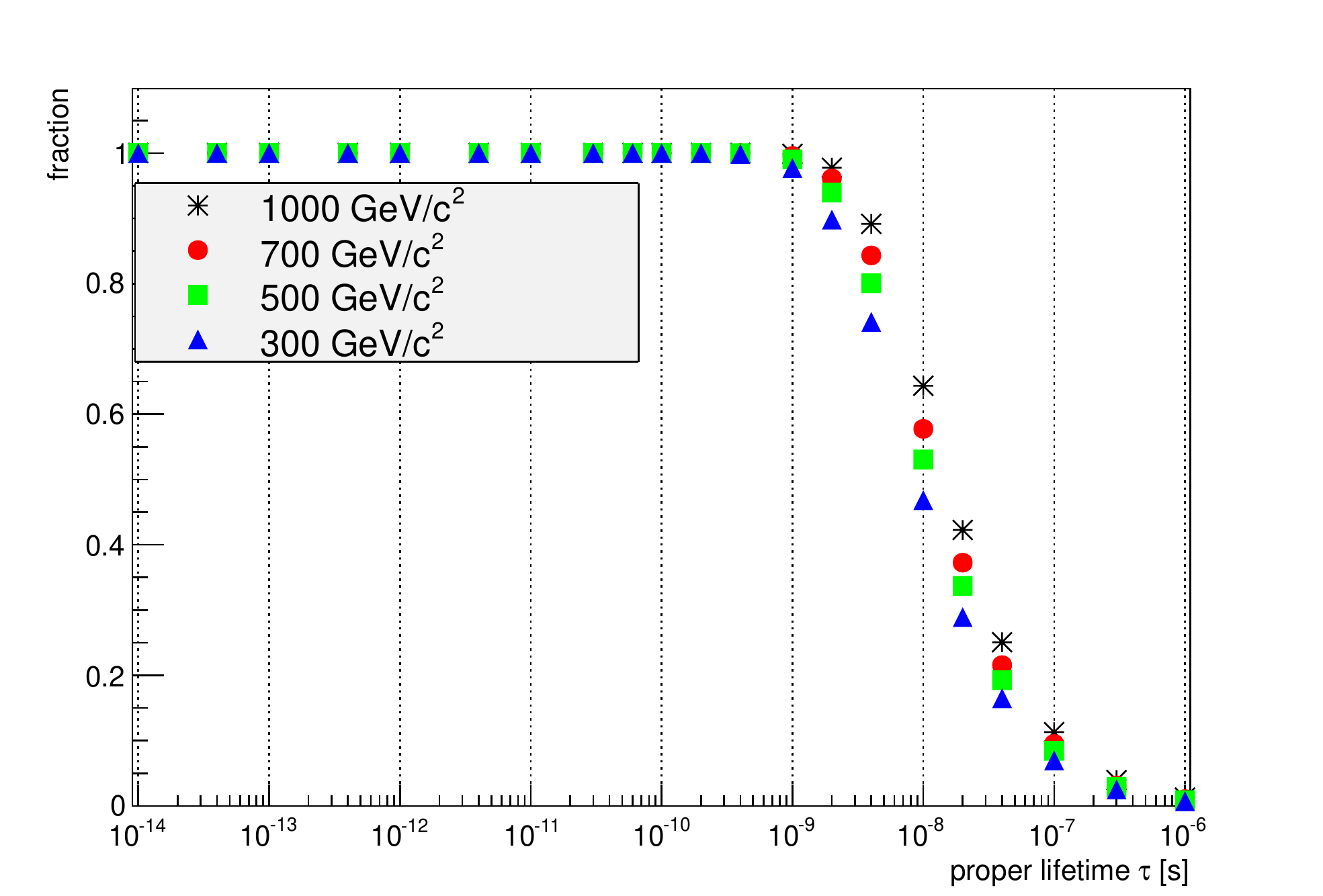}
\end{center}
\caption{Fraction of rejected decays of heavy quarks within the volume of the CMS tracking detector as a function of the proper lifetime $\tau$.}
\label{fig:TprimeSelEffTracker}
\end{figure}

Tables \ref{tab:TprimeSelEffTrackerTable} and Figure \ref{fig:TprimeSelEffTracker} summarise our estimations for a few benchmark points and as a function of the proper lifetime. The transition between the two extrema of being fully efficient and losing all decays happens within two orders of magnitudes of the lifetime between $10^{-9}$ s and $10^{-7}$ s.
\begin{table}[htbp]
\begin{tabular}{lccc}
\hline\hline
Lifetime & $2\cdot 10^{-9}$ s & $10^{-8}$ s &  $10^{-7}$ s \\ \hline
$m=300$ GeV/$c^{2}$ & 90\% & 47\% & 7\% \\ 
$m=500$ GeV/$c^{2}$ & 94\% & 53\% & 8\% \\ 
$m=700$ GeV/$c^{2}$ & 96\% & 58\% & 10\% \\ 
$m=1000$ GeV/$c^{2}$ & 98\% & 64\% & 11\% \\ \hline\hline
\end{tabular}
\caption{Fraction of rejected decays of heavy quarks within the geometrical acceptance of the CMS tracking detector. Decays within the tracker are rejected because the particle's trajectory does not reach the outermost tracker layers.}
\label{tab:TprimeSelEffTrackerTable}
\end{table}
From these results we conclude  that  heavy quark masses reaching $m_{Q}=800$ GeV/$c^2$  can be excluded at 95\% C.L. for lifetimes longer than $10^{-7}$ s. As the selection efficiency of the HSCP searches is  about 50\% for a lifetime of $\tau = 10^{-8}$~s, the published limit by the HSCP searches weakens by a factor of two, which is still sufficient to exclude the 800 GeV/$c^2$. The efficiency drops quickly for even shorter lifetimes. 

To conclude this section, we would like to mention one additional very interesting experimental strategy. As discussed in detail in \cite{PhysRevD.86.034020} heavy metastable particles may be stopped in the detectors and give out-of-time decays. This kind of signature  has been investigated by the CMS  collaboration \cite{:2012yg} using  gaps of no collisions between the proton bunches.  Particles of extremely long lifetimes (between 10~$\mu$s and 1000 s) may produce   jets which can be  triggered by the calorimeters during periods of no proton collisions. Using a data set in which CMS recorded an integrated luminosity of 4~fb$^{-1}$ and a search interval corresponding to 246 hours of trigger live time, 12 events have been observed with a background prediction of $8.6  \pm 2.4$ events. This result has been interpreted in the context of long-lived gluinos (R-hadrons) and scalar top quarks. The best limit is obtained for lifetimes larger than $10^{-6}$~s where the upper limit on the production of stop quarks has been found to be $\sigma(pp\to \tilde{t}\tilde{t}) \cdot BR(\tilde{t} \to t\tilde{\chi}^0) > 0.7$~pb. When interpreting this for pair production of heavy quarks, the cross-section of 0.7~pb corresponds  to $m_Q = 445$ GeV/$c^2$ \cite{Cacciari:2011hy}. These results are clearly less stringent than those of the stable HSCP searches discussed above, however they are able to probe signatures which can potentially escape the HSCP searches in case the particle propagates with an extremely low velocity or does not reach the outermost tracker layers before it stops.

\subsubsection{Discussion of the results \label{sec:ExpDiscussion}}
In this section we summarise  our findings  from Sections \ref{sec:ExpDirect} to \ref{sec:LongLIvedSearchesReview}, followed by a discussion of possible shortcomings and alternative search strategies. 

The following list gives an overview of the different types of searches for heavy  or long-lived particles along with the results of our reinterpretations:
\begin{itemize}
\item Prompt decays of heavy quarks: by far, the largest number of published searches for heavy quarks assume their prompt production  and decay at the proton-proton interaction vertex. We estimated that this assumption is valid for lifetimes up to $\tau < 10^{-10}$~s (c$\tau < 3$~cm). The selection efficiency decreases  for longer lifetimes and drops to 10\% - 20\% for $\tau > 10^{-9}$~s (c$\tau > 30$~cm).
\item Displaced vertices:   our reinterpretation of the ATLAS results \cite{Aad2013280} for $tW$ final states rules out cross-sections of  $\sigma \cdot BR(Q\to tW) > 0.05$~pb for lifetimes of $\rm 10^{-11}$~s  $< \tau < 3\cdot10^{-10}$~s (3~mm $< {\rm c}\tau < 10$~cm), corresponding to an excluded heavy quark mass of $m_Q < 650$~GeV/$c^2$.
\item Heavy Stable Charged Particles:  these searches require the heavy state to propagate through the full tracking detectors to be detected. Our reinterpretation of the CMS search \cite{Chatrchyan:1445273}, assuming similar behaviour as for scalar long-lived top quarks, excludes cross-sections of $\sigma>3$~fb for very long lifetimes $\tau > 10^{-7}$~s (c$\tau> 30$~m). The mass limit $m_Q < 800$~GeV/$c^2$ is still valid for lifetimes $\tau > 10^{-8}$~s (c$\tau> 3$~m).
\end{itemize}

%On the other hand, searches for HSCP can access  lifetimes larger than $ 2\cdot10^{-9}$ s for the same mass region. This means that the small window in between these two extrema, corresponding to the region (\emph{iii}) discussed in Section \ref{sec:sigDisplacedVert}, cannot be excluded yet. In principle, this invisible region could be probed by displaced vertices analyses, but the published analyses appear to not be sensitive enough yet for the heavy quark case. As mentioned in Section \ref{sec:displaced}, displacements beyond $50$ cm cannot be detected so that these searches are not able to cover the full tracker volume. 

\begin{figure}[htbp]
\begin{center}
\includegraphics[width=0.52\textwidth]{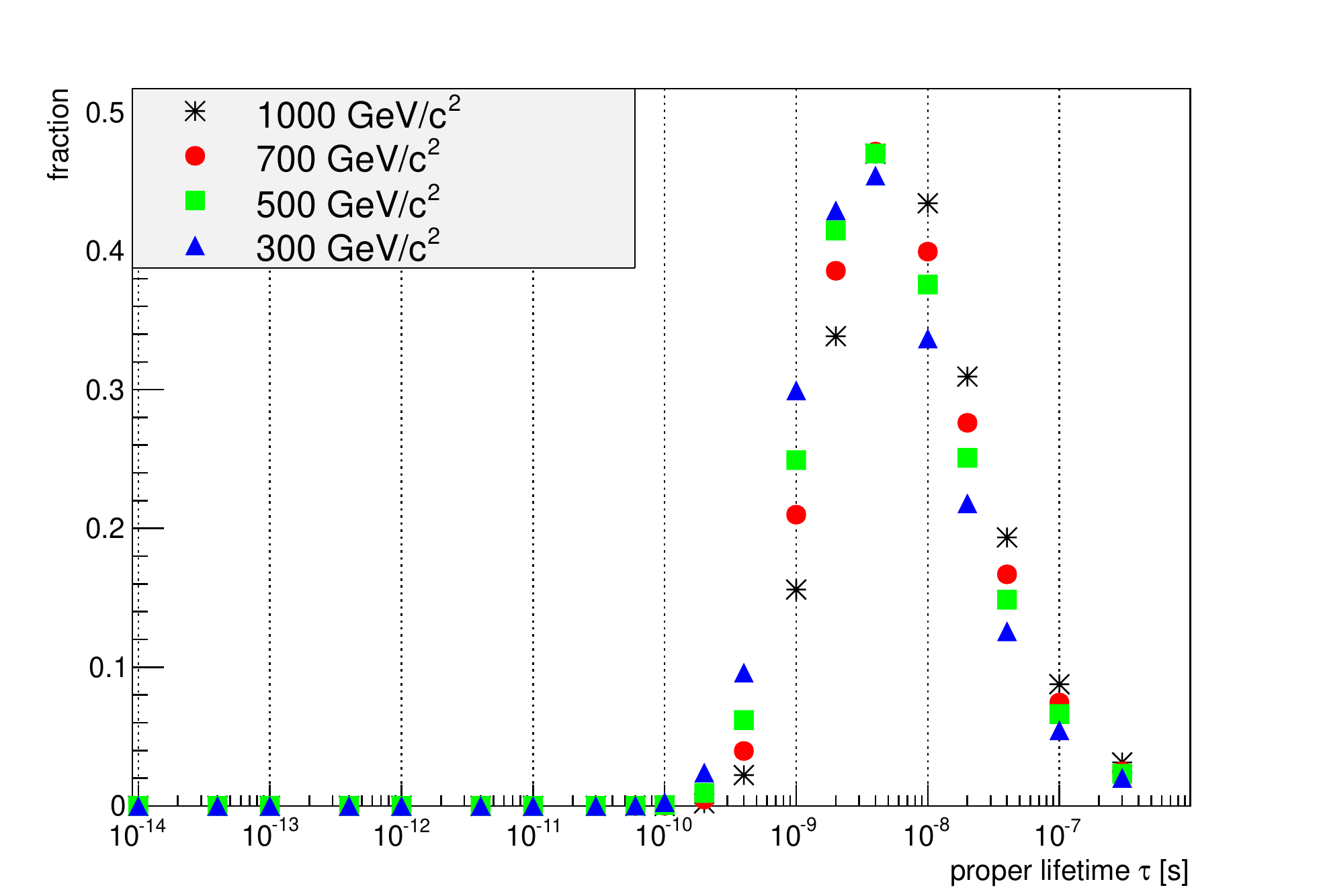}
\end{center}
\caption{Fraction of decays of heavy quarks between the acceptance of  the displaced vertex analysis and HSCP analyses.}
\label{fig:TprimeSelBetween}
\end{figure}

This is obviously still  a patchwork of different sensitivities for different lifetimes and different models and methods. We also see  that  a small geometrical window of inefficiency remains. This is the region beyond displacements of more than about 50 cm  but within the volume of the tracking devices.  In this region, both the searches for displaced vertices and the searches for HSCP do not reconstruct any events.  The exact size of this geometrical region depends on the experiment. We estimate the fraction of decays of heavy quarks within this geometrical region based on our simulation from the previous sections.   The result is shown in Figure \ref{fig:TprimeSelBetween}. We see that in the worst case scenario only 50\% of the decays are lost  for a lifetime of $\tau = 3\cdot 10^{-9}$ s. This represents a moderate decrease of the selection efficiency for a narrow window, so that we can expect a full coverage of the lifetime spectrum in the future.

Yet, some of the assumptions made in this section may have certain shortcomings. First of all, we considered a general framework in which the new heavy quarks are pair-produced, with kinematics independent of the lifetime of the particle. Our motivation was that the current searches for long-lived R-hadrons formed from stable stops generally assume that strong pair production dominates. In supersymmetric models, however, the situation changes drastically when the squark masses become closer to the TeV scale. As pointed out in \cite{Johansen:2010ac}, the squark pair production from light quarks can dominate the gluon fusion mechanism for large masses due to the decreasing probability for the gluons to carry large fractions of the proton momentum. For scenarios where either the squark mass or the mass of the gluino propagator reaches $1$ TeV, Higgs and squark weak production start to dominate. The production channels then become more and more model-dependent for increasing masses, arguing for dedicated MC simulations above such mass values. %Because of the lower production cross-section, models involving such new heavy quarks could be very challenging to detect at the LHC, depending on how well the signal signatures can be discriminated.

Secondly, we assumed that the heavy quark pair production cross-sections were of the same magnitude as the rate of particles hadronising into bound states, while we stressed in Section \ref{sec:HSCPsignatures} that this is not the case in general. In particular, the $\eta_{Q}$ quarkonium production rate is expected to be smaller than the $Q\bar{Q}$ pair production cross-section, due to the possibly large suppression arising from the bound state wave function, which also leads to a different decay phenomenology. Precise calculations would be needed to refine our analysis and the above discussion might need to be adjusted in case of strongly bound states.

As a possible suggestion to extend the reach of the above analyses, another search topology may be the case in which a heavy quark decays within the tracker volume. The signature to be searched for may then be one or more displaced jets, which trigger the event, and originate from the heavy quark decay. In addition,  a short track with a few hits and large d$E$/d$x$ pointing to the decay vertex could be required. This signature may have relatively low backgrounds as the jets should have a large transverse momentum. The tracks within these jets are not required to arise from the primary interaction vertex which makes track reconstruction difficult. To circumvent this, one could perform a regional search for short tracklets pointing towards the energy deposit in the calorimeter.

%While precise calculations would be needed to refine our analysis, the above discussion may thus need to be %adjusted in case of strongly bound states. 

%\subsubsection{Alternative search topologies \label{sec:NewSearchTopologies}}

\section{Summary and conclusions \label{sec:conclusion}}

The current searches for heavy quarks at the LHC mostly ignore the option of long lifetimes. Although a new chiral family of fermions is now strongly disfavoured from the recent Higgs search results, such a scenario remains of prime importance when considering vector-like quarks beyond the Standard Model. As non-chiral fermions decouple in the limit of vanishing mixing with the lighter generations, they can be long-lived and form bound states. Dedicated searches for novel heavy $Q\bar{Q}$ quarkonia, $Qqq$ baryons and $Q\bar{q}$\ ($\bar{Q}q$) open-flavour mesons might provide promising strategies for future investigations.\

As we have detailed, our reinterpretation of the HSCP searches indicates that quark masses of $800$ GeV/$c^2$ can be excluded at $95\%$ CL for lifetimes longer than $10^{-8}$ s. This result however assumes that the interactions of the heavy quarks with the detector material are similar to that of stop particles. As discussed in this review, there are potential restrictions to such an assumption, and the forthcoming searches should be interpreted in a model-specific context. Additionally, it is usually considered that the bound states production cross-sections are comparable to that of pair-produced heavy quarks. This is not necessarily the case for the reasons we presented. 

Finally, searches for prompt decays are still valid for long lifetimes to a certain extent, but the assumptions about branching ratios and the potential loss of efficiency need to be considered. We showed that these analyses start to become insensitive for lifetimes longer than $10^{-10}$ s, corresponding to $Q-q$ quark couplings below the $10^{-9}$ level. For lifetimes beyond $10^{-10}$ s, searches for displaced vertices in the tracker volume have been shown to play an important role.

\textbf{Acknowledgements}
The work of M.B. is supported by the National Fund for Scientific Research (F.R.S.-FNRS, Belgium) under a FRIA grant. A.S. is supported by the German Research Foundation (DFG) under an Emmy Noether grant. The authors would like to thank Daniel Stolarski and Yevgeny Kats for their comments on the early version of this manuscript. We acknowledge the organisers of the "Focus Workshop on Heavy Quarks at LHC" in Taipei (NTU) for the discussions that initiated this work.

\bibliography{690254_BuchkremerSchmidt}
%\bibliographystyle{plain}
%\bibliography{Template7bis}
%\printbibliography
% Produces the bibliography via BibTeX.

\end{document}